\documentclass[aps,onecolumn,pre,showpacs,groupedaddress,showkeys,amsmath,amssymb,superscriptaddress]{revtex4}

\usepackage{float}

\usepackage{amsmath}
\usepackage{amssymb}
\usepackage{graphicx}
\usepackage{bm}
\usepackage{epic}
\usepackage{eepic}
\usepackage{pifont}
\usepackage[utf8]{inputenc}
\usepackage{rotating}
\usepackage{color}
\usepackage{nicefrac}
\usepackage{ulem}
\usepackage[caption=false]{subfig}
\usepackage{array}

\begin{document}
\draft
\title{Unsteady mass transfer from a core-shell cylinder in crossflow}
\author{Clément Bielinski}
\author{Nam Le}
\author{Badr Kaoui}
\email{badr.kaoui@utc.fr}
\affiliation{Biomechanics and Bioengineering Laboratory,\\
Université de technologie de Compiègne, CNRS,\\
60200 Compiègne, France}
\date{\today}
\begin{abstract}
Mass transfer from a composite cylinder - made of an inner core and an outer enveloping semipermeable shell - under channel crossflow is studied numerically using two-dimensional lattice-Boltzmann simulations.
The core is initially loaded with a solute that diffuses passively through the shell towards the fluid.
The cylinder internal structure and the initial condition considered in this study differ and thus complement the classical studies dealing with homogeneous uncoated cylinders whose surfaces are sustained at either constant concentration or constant mass flux.
Here, the cylinder acts as a reservoir endowed with a shell that controls the leakage rate of the encapsulated solute.
The transition from steady to unsteady laminar flow regime, around the cylinder, alters the released solute spatial distribution and the mass transfer efficiency, which is characterized by the Sherwood number (the dimensionless mass transfer coefficient).
Moreover, the reservoir involves unsteady and continuous boundary conditions, which lead to unsteady and nonuniform distribution of both the concentration and the mass flux at the cylinder surface.
The effect of adding a coating shell is highlighted, for a given ratio of the cylinder diameter to the channel width, by extracting a correlation from the computed data set.
This new correlation shows explicit dependency of the Sherwood number upon the shell solute permeability (the shell mass transfer coefficient). 
\end{abstract}
\maketitle
\section{Problem statement}
\label{sec:intro}
Mass transfer from or to particles is of fundamental and practical interest. 
It is encountered in many natural phenomena and in industrial processes.
For example, in the cardiovascular system where red blood cells transport oxygen and carbon dioxide, or in fluidized bed reactors largely used in chemical engineering.
In such systems, mass transfer mainly takes place at the fluid adjacent to the surface of the particles.
It might be limited by factors, such as particle dissolution or chemical reactions.
On the contrary, imposing an external fluid flow counterbalances these limiting factors, and it enhances mass transfer due to forced convection.
Mass transfer can also be controlled by covering the particles with a shell, or a membrane, that tends to slow down the mass flux rate.
The objective of this study is to investigate mass transfer under crossflow from a circular cylinder having two specific characteristics: 1 - being a reservoir that means setting a nonsustained initial condition, and that involves unsteady continuous boundary conditions, and 2 - being coated with a semipermeable shell that adds a resistance to mass transport from the reservoir to the fluid. 

Many classical studies have characterized the dependency of the mass transfer coefficient on the speed of the flow, shape of the particles and on the transport properties of both the solute and the solvent, as done in Ref.~\cite{Whitaker1972}.
Empirical correlations have been proposed to give the Sherwood number (the dimensionless mass transfer coefficient) as a function of mainly the Reynolds and the Schmidt numbers ${\rm Sh}=f\left( {\rm Re},{\rm Sc}\right)$; however, the effect of having a shell has not been characterized.
On the one hand, most conducted studies under flow conditions deal with rather naked particles, \textit{i.e.}, devoid of any shell \cite{Clift1978,Michaelides2006,Incropera2017}.
On the other hand, existing studies that describe mass transfer from particles with a shell are valid only under static conditions, \textit{e.g.}, fluid at rest \cite{Crank1975,Siepmann2012,Romano2017}.
Moreover, the surface of the particles in these studies is maintained at either constant Dirichlet (concentration) or Neumann (mass flux) boundary conditions, which do not apply for the reservoirs whose surface develops unsteady boundary conditions.
This study complements the previous works by examining for the first time how the flow, in presence of a shell, alters mass transfer from a cylindrical reservoir made of an inner core - initially loaded with a solute - and an outer enveloping shell; see Fig.~\ref{fig:geometry}.
It examines how the reservoir - through its initial condition that decays with time and its continuous concentration and mass flux on the surface - changes the mass transfer mechanism compared to the classical studies that consider a homogeneous cylinder with constant boundary conditions \cite{Krall1973,Karniadakis1988,Haeri2013}. 

The solute leaks outside the reservoir by passively diffusing through the shell, and then being transported by diffusion-advection in the outer passing fluid.
Here, the shell is considered to be semipermeable, \textit{i.e.}, it is permeable to the solute, but not to the solvent.
Because of this, the fluid cannot penetrate into the reservoir; and therefore, the reservoir is seen as a nonporous solid obstacle by the fluid flow.
The hydrodynamic part of the problem corresponds to the classical extensively studied problem of flow past a circular cylinder \cite{Shair1963,Chen1995,Sahin2004}.
The mass transfer part and its one-way coupling with the hydrodynamic, however, is not explored so far to the best knowledge of the authors.
Solving solute transport in such composite medium presents extra mathematical as well as numerical challenges.
This is an initial-value problem, for which three advection-diffusion equations need to be solved in each domain, the core, the shell and the fluid, while considering unsteady continuous mass transfer boundary conditions at both the core-shell and the shell-fluid interfaces. 
A two-dimensional numerical approach, based on a two-component lattice Boltzmann method (LBM), has been proposed by one of the authors to study flow and mass transfer around a single confined core-shell cylinder by decomposing the computational domain into subdomains \cite{Kaoui2017}. 
In this previous study, mass transfer is reported as a function of the P\'{e}clet number, at a fixed Reynolds number.
Here, a wide range of the Reynolds number is explored in order to examine how the transition of the laminar flow from steady to unsteady regime alters the mass transfer from a cylindrical core-shell reservoir. 
The effect of having a coating semipermeable shell, on mass transfer under crossflow condition, is studied systematically and quantified by a correlation.

This article is organized as follows. 
In Sec.~\ref{sec:numerical_method}, the numerical method is given and explained.
In Sec.~\ref{sec:results_discussions} the results are reported and discussed. 
An emphasis is given to analyze local and global mass transfer quantities.
The effect of the solute permeability is investigated and accordingly a new mass transfer correlation is proposed.
Discussion within the context of existing classical correlations, obtained for cylinders without a shell and whose surface concentration is constant \cite{Hilpert1933,Churchill1977}, is provided.
Conclusions are given in Sec.~\ref{sec:conclusion}.
\section{Problem setup and computational method}
\label{sec:numerical_method}
\begin{figure}[b]
\centering
\includegraphics*[width = 0.8\textwidth]{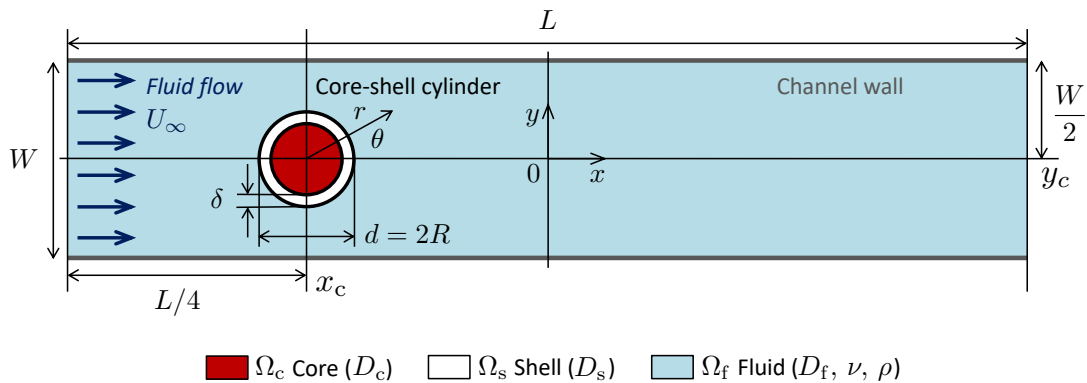}
\caption{Geometry of the problem.
The solute-free fluid enters the channel from the left with uniform velocity profile, passes the stationary core-shell cylinder, and exist the channel with constant pressure.
The domain is decomposed into three adjacent subdomains: the core, the shell and the fluid.
The core is initially loaded with a solute that leaks out, through the semipermeable shell, toward the outer passing fluid.
}
\label{fig:geometry}
\end{figure}
In absence of turbulence, at small Reynolds numbers ${\rm Re} < 200$ \cite{Haeri2013}, the problem of flow perpendicular to an infinite circular cylinder can be solved in two-dimensional space.
The characteristics of the numerical setup are shown in Fig.~\ref{fig:geometry}.
The cylinder has an outer radius $R$, and it is composed of a core coated with a semipermeable shell of uniform thickness $\delta$.
This latter is hold constant at $\delta=0.3R$ along this work.
Both the core and the shell have the same central axis located at $(x_c,y_c)$.
They are placed in a channel, with length $L=20R$ and width $W=4R$, at a distance of $x_c = L/4$ from the channel inlet and at $y_c = 0$ (channel centerline).
The solvent, here a fluid, enters the channel from the left with a uniform flow velocity $U_{\infty}$ and with zero solute concentration $c_{\infty} = 0$, and exits the channel at the right with constant pressure and zero-gradient of the concentration.
The reservoir is stationary and the hydrodynamic part of the problem corresponds to the classical case study of flow past a stationary circular cylinder.
The flow does not make any distinction between the core and the shell. 
It just sees a nonporous solid obstacle of radius $R$ regardless of the internal structure.
For the hydrodynamic part, the computation domain consists of only two subdomains: the fluid and the obstacle, \textit{i.e.}, the reservoir (the core and the shell all together).
However, for the mass transfer part, the domain is decomposed into three adjacent subdomains that correspond to the core $\Omega_{\rm c}$, the shell $\Omega_{\rm s}$ and the fluid $\Omega_{\rm f}$, where the solute has different diffusion coefficients ~\cite{Morvan1989,Kaoui2017}: $D_{\rm c}$ in the core, $D_{\rm s}$ in the shell, and $D_{\rm f}$ in the fluid.
Both the core and the shell are homogeneous media.
The shell is permeable only to the solute, but not to the solvent, \textit{i.e.}, it is impermeable to the fluid. 
The shell separates and moderates solute mass transfer between the core (the donor compartment) and the fluid (the receptor medium), with $D_{\rm c}=D_{\rm f}$, as if the core is filled with the same solvent as outside the reservoir.
Moreover, the shell is considered to be a solid medium with $0.1 \leq D_{\rm s}/D_{\rm f} \leq 1$.
\subsection{Mathematical formulation}
\label{subsec:mathematical_formulation}
\textit{Governing equations - }
The problem requires solving the Navier-Stokes equations~(\ref{eq:NS1}) and (\ref{eq:NS2}) for the fluid, which is considered to be incompressible Newtonian fluid, and the advection-diffusion equation (\ref{eq:adv_diff}) for solute transport:
\begin{equation}
\frac{\partial {\bf u}}{\partial t} + {\bf u} \cdot \nabla {\bf u} = -\frac{\nabla p}{\rho} + \nu \nabla ^2 {\bf u},
\label{eq:NS1}
\end{equation}
\begin{equation}
\nabla \cdot {\bf u} = 0,
\label{eq:NS2}
\end{equation}
\begin{equation}
\frac{\partial c}{\partial t} + {\bf u} \cdot \nabla c = \nabla \cdot \left(D \nabla c\right)\text{,}
\label{eq:adv_diff}
\end{equation}
where ${\bf u}(x,y,t)$ is the local solvent velocity, and $p(x,y,t)$ the local pressure at time $t$.
$\rho$ and $\nu$ are the mass density and the kinematic viscosity of the fluid, respectively. 
$c(x,y,t)$ is the local concentration of the solute, and $D(x,y)$ the local diffusion coefficient of the solute,
\begin{equation}
D(x,y)=\begin{cases}
D_{\rm c} & \text{if $(x,y) \in \Omega_{\rm c}$},\\
D_{\rm s} & \text{if $(x,y) \in \Omega_{\rm s}$},\\
D_{\rm f} & \text{if $(x,y) \in \Omega_{\rm f}$}.
\end{cases}
\end{equation}
Equation~(\ref{eq:adv_diff}) is solved without the advection term in the cylindrical reservoir that is stationary, and where ${\bf u}(x,y,t)=0$ in both the core and the shell .

\textit{Initial conditions - }
The nonsteady and nonlinear coupled partial differential equations~(\ref{eq:NS1}), (\ref{eq:NS2}) and (\ref{eq:adv_diff}) are solved for the initial condition,
\begin{equation}
c(x,y,t=0)=\begin{cases}
c_0 & \text{if $(x,y) \in \Omega_{\rm c}$},\\
0 & \text{elsewhere},
\end{cases}
\label{eq:ic}
\end{equation}
that models a reservoir, with an initially loaded solute concentration $c_0$ uniformly distributed within the core.
The initially established concentration gradient induces solute transport from the core to the fluid (out-of-equilibrium state), which leads to concentration decay over time inside the reservoir until exhaustion of the solute (equilibrium state).
There is no sustained supply of solute that would maintain the initial concentration, and a constant concentration at the reservoir surface.

\textit{Boundary conditions - } 
For the flow part, nonslip velocity boundary condition ${\bf u}=(0,0)$ is set on the channel walls $y=\pm W/2$ and on the outer surface of the reservoir $r=R$.
$r$ is the distance from the center of the reservoir $(x_{\rm c},y_{\rm c})$, see Fig.~\ref{fig:geometry}.
The flow is uniform and parallel to the channel walls at the inlet ${\bf u}(-L/2,y)=(U_{\infty},0)$, and it exits the channel with constant pressure.
For the mass transfer part, zero mass flux Neumann boundary condition $\left. \partial c / \partial y \right|_{y=\pm W/2}=0$ is set on the channel walls (\textit{i.e.} impermeable walls), zero concentration Dirichlet boundary condition $c_{\infty}=0$ is set at the channel inlet, and zero gradient of the concentration $\left. \partial c / \partial x \right|_{x=+L/2}=0$ is set at the channel outlet.
Neither the concentration nor the mass flux are constant at the surface of the reservoir, they are rather time-varying quantities that evolve in time when the reservoir is loosing its initial solute load.
Thus, in absence of any interfacial resistance, the continuity of both the concentration and the mass flux emerges at the core-shell interface ($r=R-\delta$),
\begin{equation}
c_{\rm core} = c_{\rm shell}
\quad
\text{and}
\quad
\left.D_{\rm c}\frac{\partial c}{\partial r}\right|_{\rm core} = \left.D_{\rm s}\frac{\partial c}{\partial r}\right|_{\rm shell},
\label{eq:bc_cs}
\end{equation}
and at the shell-fluid interface ($r=R$):
\begin{equation}
c_{\rm shell} = c_{\rm fluid}
\quad
\text{and}
\quad 
\left.D_{\rm s}\frac{\partial c}{\partial r}\right|_{\rm shell} = \left.D_{\rm f}\frac{\partial c}{\partial r}\right|_{\rm fluid},
\label{eq:bc_sf}
\end{equation}
where the subscripts stand for quantities evaluated on the sides facing the core, the shell and the fluid.
These unsteady continuous boundary conditions lead to different mass transfer scenario compared to the largely used constant boundary conditions, and to nonuniform boundary conditions examined recently in Ref.~\cite{Vandadi2016}.
\subsection{Key physical quantities}
Mass transfer from a cylindrical core-shell reservoir is expected to depend on the Reynolds and the Schmidt numbers $(\rm{Re},\rm{Sc})$, as is the case for other type of particles, and in particular on its shell properties. 
The flow pattern that develops around the cylindrical reservoir is dictated by the imposed Reynolds number that measures the importance of inertia with respect to viscous forces,
\begin{equation}
\mathrm{Re} = \frac{d\,U_{\rm max}}{\nu},
\end{equation}
where $d = 2R$ is the reservoir diameter, and $U_{\rm max}=3U_{\infty}/2$ the maximum velocity at the channel centerline in absence of any obstacle.
Because the cylinder is confined between two walls, $U_{\rm max}$ is taken as the characteristic velocity, instead of $U_{\infty}$, which is usually used for unbounded geometries.
In this study, $0.01 \leq {\rm Re} \leq 180$.
The flow pattern is also influenced by the position of the cylinder with respect to the channel inlet, here $x_c=L/4$, and on the blockage degree that is defined as the ratio of the cylinder outer diameter to the channel width $\mathrm{B} = d/W$ \cite{Sahin2004}.
This is set to $\mathrm{B} = 0.5$ to model situation of moderate effect of confinement.
Moreover, the mass transfer depends also on the Schmidt number that is defined as the ratio between the rates of momentum and mass diffusion,
\begin{equation}
\mathrm{Sc} = \frac{\nu}{D_{\rm f}}.
\end{equation}
It is typically of order unity for diffusion in gases, and is larger for diffusion in liquids.
Here, $1 \leq {\rm Sc} \leq 25$. 
For the core-shell reservoir case, the mass transfer is expected to depend also on the shell permeability to solute (the shell mass transfer coefficient), which is defined as the ratio of the solute diffusivity in the shell $D_{\rm s}$ to the thickness of the shell $\delta$, $p_{\rm s} = D_{\rm s}/\delta$.
$p_{\rm s}$ measures how easily the solute diffuses through the shell. 
Hereafter, it is scaled as,
\begin{equation}
{\rm P} = \frac{d}{D_{\rm f}}p_{\rm s},
\label{eq:scaled_p}
\end{equation}
and varied within the range $0.66 \leq {\rm P} \leq 6.67$.
Five observable quantities are measured and analyzed to characterize mass transfer from a cylindrical core-shell reservoir:
\begin{enumerate}
\item The solute concentration $c_{\rm s}(\theta,{\rm T})$ at the surface of the reservoir $(r=R)$, which is computed using bilinear interpolation of the known concentration at the on-lattice points.
This necessitates high grid resolution in order to achieve smooth concentration interpolation.
The angular position $\theta$ is defined in Fig.~\ref{fig:geometry} with $\theta = 0$ at the rear of the particle (downstream) and $\theta = \pi$ at the front (upstream).
$\mathrm{T} = (D_{\rm f}/d^2)t$ is the dimensionless time,
\item The surface mass flux,
\begin{equation}
\varphi(\theta,{\rm T}) = -D_{\rm f}\left[\frac{\partial c}{\partial r}\right]_{r=R},
\label{eq:local_flux}
\end{equation}
\item The local Sherwood number that quantifies the ratio of the convective mass transfer to the rate of diffusive mass transport at the surface of the reservoir,
\begin{equation}
\mathrm{Sh}(\theta,{\rm T}) =  \frac{d}{D_{\rm f}}h(\theta,{\rm T}) = \frac{d}{\delta_{\rm BL}(\theta,{\rm T})},
\label{eq:local_sh}
\end{equation}
where $h(\theta,{\rm T}) = \varphi(\theta,{\rm T})/\left[c_{\rm s}(\theta,{\rm T}) - c_{\infty}\right]$
is the convective mass transfer coefficient, and $c_{\infty}$ the concentration at the channel inlet. 
$\mathrm{Sh}(\theta,{\rm T})$ is inversely proportional to the concentration boundary layer thickness $\delta_{\rm BL}(\theta,{\rm T})$,
\item The average Sherwood number, which is computed as the integration of the local Sherwood number Eq.~(\ref{eq:local_sh}) over the surface of the reservoir \cite{Kaoui2017},
\begin{equation}
{\mathrm{Sh}}({\rm T}) = \frac{1}{2\pi R}\int_{\theta=0}^{2\pi}\mathrm{Sh}(\theta,{\rm T})ds(\theta),
\label{eq:average_sh}
\end{equation}
and where $s$ is the curvilinear coordinate along the cylinder circumference,
\item The steady average Sherwood number, denoted by ${\rm Sh}$ (without ${\rm T}$ between parentheses), is used hereafter to derive a mass transfer correlation.
\end{enumerate}
\subsection{Lattice-Boltzmann method}
\label{subsec:lbm}
In this study, the lattice Boltzmann method (LBM) is opted to compute both the flow and the mass transfer \cite{Wolf-Gladrow2000,Succi2001,Sukop2006,Mohamad2011,Kruger2016}, instead of solving Eqs.~(\ref{eq:NS1}), (\ref{eq:NS2}), and (\ref{eq:adv_diff}), as done previously by one of the authors in Refs.~ \cite{Ferrari2012,Kaoui2017,Kaoui2018,Kaoui2020}.
In the limit of small Mach numbers (ratio of the speed of a fluid particle in a medium to the speed of sound in that medium) and Knudsen numbers (ratio of the molecular mean free path to
the macroscopic characteristic length scale) the LBM computes with good approximation the flow of an incompressible Newtonian fluid.

Two distribution functions $f_i$ and $g_i$ are needed with their respective lattice Boltzmann equations.
The $f_i$ is used for the flow part, and it gives the probability to find a population of fluid particles at the discrete vector position ${\bf r} = (x,y)$ at time $t$, with the $i$th discrete velocity vector ${\bf e_i}$, where $i$ depends on the lattice type.
Here, the D2Q9 lattice is used, with $i=0-8$. 
The nine distribution functions $f_i$ are computed, at each node and at each time iteration, by the discrete lattice Boltzmann equation with the Bhatnagar-Gross-Krook (BGK) collision operator,
\begin{equation}
f_i({\bf r}+{\bf e}_i,t+1) - f_i({\bf r},t) = -\frac{1}{\tau_{\nu}} \left[f_i({\bf r},t) - f_i^{\rm eq}({\bf r},t)\right],
\label{eq:discrete_LBE_flow}
\end{equation}
where $f_i^{\rm eq}$ is the equilibrium distribution function given by,
\begin{equation}
f^{\rm eq} _{i}({\bf r},t)= \omega _i \rho \left[1 + 3({\bf u}\cdot {\bf e}_i) + \frac{9}{2}({\bf u}\cdot{\bf e}_i)^2 -\frac{3}{2}({\bf u})^2 \right].
\end{equation}
The weight factors $\omega_i$ for the D2Q9 lattice are: $\omega_i=\frac{4}{9}$ for $i=0$, $\omega_i=\frac{1}{9}$ for $i=1-4$, and $\omega_i=\frac{1}{36}$ for $i=5-8$.
The relaxation time ${\tau_{\nu }}$ is related to the fluid kinematic viscosity $\nu$ via the relationship,
\begin{equation}
\nu = \frac{1}{3}(\tau_{\nu} - \frac{1}{2}).
\label{eq:tau_nu}
\end{equation}
The local density $\rho$ and the local macroscopic velocity ${\bf u}$ are computed respectively by the zeroth and the first order moments of the distribution function $f_i$,
\begin{equation}
\rho(x,y,t) = \sum _{i=0}^{8} f_i (x,y,t) \quad {\rm and}\quad {\bf u}(x,y,t) = \frac{1}{\rho}\sum _{i=0}^{8} f_i (x,y,t){\bf e}_i.
\end{equation}
The zero non-slip conditions are set on the surface of the cylindrical reservoir and on the channel walls using the bounce-back boundary condition.
Zou and He boundary conditions \cite{Zou_He1997} are used to set uniform incoming flow ${\bf u} = (U_{\infty},0)$ at the channel inlet, and constant fluid density at the outlet to recover constant pressure outflow boundary condition.

The distribution function $g_i$ is used for the mass transfer solver with its corresponding discrete lattice Boltzmann equation,
\begin{equation}
g_i({\bf r}+{\bf e}_i,t+1) - g_i({\bf r},t) = -\frac{1}{\tau_{D}} \left[g_i({\bf r},t) - g_i^{\rm eq}({\bf r},t)\right].
\label{eq:discrete_LBE_mass}
\end{equation}
$g_i$ exists on the same D2Q9 lattice as $f_i$, and it has the equilibrium distribution function,

\begin{equation}
g^{\rm eq} _{i}({\bf r},t)= \omega _i c \left[1 + 3({\bf u}\cdot {\bf e}_i) + \frac{9}{2}({\bf u}\cdot{\bf e}_i)^2 -\frac{3}{2}({\bf u})^2 \right],
\end{equation}
where the weight factors $\omega_i$ are the same as those of the flow solver, and the relaxation time ${\tau_D}$ is related to the diffusion coefficient $D$,
\begin{equation}
D = \frac{1}{3}(\tau_{D} - \frac{1}{2}).
\label{eq:tau_d}
\end{equation}
Different relaxation times are used to set different diffusion coefficients is each computational subdomain, $\tau_{D_{\rm c}}$ in the core, $\tau_{D_{\rm s}}$ in the shell and $\tau_{D_{\rm f}}$ in the surrounding fluid (see Fig.~\ref{fig:geometry}).
The local concentration is computed by the zeroth-order moment of the distribution function $g_i$,
\begin{equation}
c(x,y,t) = \sum _{i=0}^{8} g_i(x,y,t).
\end{equation}
The zero mass flux is set on the channel walls using the standard bounce-back boundary condition \cite{Sukop2006}.
The zero concentration at the inlet, and the zero concentration gradient at the outlet are set using the boundary conditions developed by Inamuro \textit{et al.} for thermal flows \cite{Inamuro2002} due to the temperature-concentration analogy.

The numerical method has been successfully benchmarked with the limiting case studies of Poiseuille flow, pure diffusion of a Gaussian hill, advection-diffusion of a Gaussian hill by a uniform flow, and flow past a circular cylinder.
Similar validations can be found, for example, in Refs.~\cite{Sukop2006,Kruger2016}, and thus, are not reported here.
For the presently studied situation, mass transfer from a core-shell reservoir in crossflow, there is no available analytical or numerical solutions counterpart for validation.
A systematic study of the results dependency on the grid resolution has been necessary, and it is found that high resolution simulations lead to stable and reliable simulations.
For larger grid resolutions, the Zou and He boundary conditions \cite{Zou_He1997} computes correctly the expected flow pattern. 
The stairslike approximation of the cylinder surface inherited from the bounce-back boundary condition is smooth enough, and the wake as well as the vortices behind the cylinder are well captured. 
Furthermore, there are enough grid points within the gap between the cylinder surface and the channel walls for the flow to be accurately resolved.
Similarly, there are enough grid points within the shell to resolve the diffusion precisely and to compute the bilinear interpolation of the surface concentration with good accuracy.
Relative deviation of the computed steady average Sherwood number $\mathrm{Sh}$, with respect to the finest tested resolution $2401\times481$, when reducing the grid resolution is reported in Table \ref{tab:mesh_dependency_Re_180} for the largest used Reynolds number (${\rm Re}=180$).
Measured deviations are below $5\%$, even when dividing the original referential resolution by half, which is satisfactory enough to trust the computed results. 
The refined grid resolution $2001\times401$ is opted to carry out all the simulations, with a reservoir of radius $R=100$ and a shell thickness of $\delta = 30$.
\begin{table}[H]
\centering
\begin{tabular}{|>{\raggedright\arraybackslash}m{25mm}|m{15mm}|m{15mm}|}
\hline
\multicolumn{1}{|>{\centering\arraybackslash}m{25mm}|}{Grid resolution} 
& \multicolumn{1}{>{\centering\arraybackslash}m{15mm}|}{$\bf{\mathrm{Sh}}$} 
& \multicolumn{1}{>{\centering\arraybackslash}m{20mm}|}{$\bf{\text{Deviation}~(\%)}$}\\
\hline
$2401\times481$ & 20.69 & -\\
\hline
$2001\times401$ & 20.54 & 0.72\\
\hline
$1601\times321$ & 20.30 & 1.88\\
\hline
$1001\times201$ & 19.87 & 3.96\\    
\hline
\end{tabular}
\caption{Grid resolution dependency of the computed steady average Sherwood number ${\rm Sh}$ at the largest used Reynolds number $\mathrm{Re} = 180$. $\mathrm{Sc} = 5$ and $\mathrm{P} = 3.70$.
}
\label{tab:mesh_dependency_Re_180}
\end{table}

\begin{figure}[H]
\centering
\includegraphics[width = 0.45\textwidth]{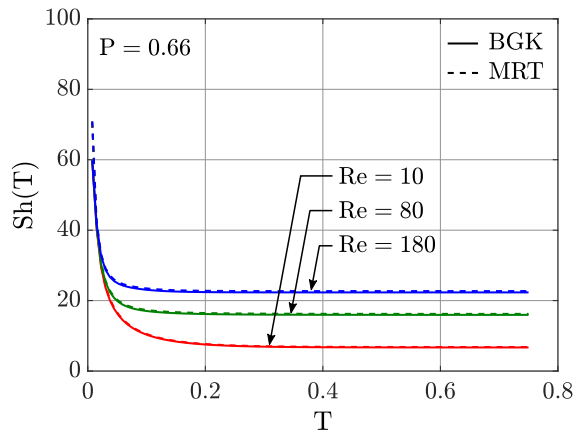}
\caption{Instantaneous average Sherwood number ${\rm Sh}({\rm T})$ computed with both the BGK and the MRT schemes for the dimensionless solute shell permeability ${\rm P} = 0.66$, which corresponds to the smallest value of the LBM relaxation time used in this study $\tau_{D_{\rm s}} = 0.509$.
The Schmidt number is ${\rm Sc}=1$.}
\label{fig:comp_BGK_MRT}
\end{figure}

The introduction of a coating shell has added additional numerical challenges to conduct stable and accurate simulations with the LBM.
Achieving desired larger Reynolds numbers, lower shell solute permeabilities and larger Schmidt numbers requires using LBM relaxation times (Eq.~(\ref{eq:tau_nu}) and (\ref{eq:tau_d})) that tend to $1/2$ and that lead to numerical instabilities. 
Simulations are performed with both the Bhatnagar-Gross-Krook (BGK) and the multiple relaxation time (MRT) collision operators \cite{Mohamad2011,Kruger2016}.
Figure~\ref{fig:comp_BGK_MRT} shows the computed Sherwood number ${\rm Sh}$ for the smallest value of the LBM relaxation time used in this study $\tau_{D_{\rm s}} = 0.509$, where both schemes give almost the same curves.
Because the MRT is time consuming, most of the simulations are conducted with the BGK scheme.

Additional simulations are conducted for a cylinder, without a shell, whose surface is maintained at  constant concentration $c_{\rm w}$ (Dirichlet boundary condition).
This allows comparison with the results of the core-shell cylinder, and thus, to appreciate the effects of the shell and the continuous boundary conditions.
The constant concentration $c_{\rm w}=1$ on the cylinder surface ($r=R$) is implemented within the lattice Boltzmann method using the recently proposed scheme of Zhang \textit{et al.} \cite{Zhang2012} for concentration, which is similar to Ladd's scheme for momentum \cite{Ladd1994} .
Here, another distribution function is used $g_{i'}\left(\bm{x}_{\rm f},t+1\right)$ on any fluid node $\bm{x}_{\rm f}$ located in the neighborhood of a solid node of the cylinder, and it is updated after the collision step as,
\begin{equation}
g_{i'}\left(\bm{x}_{\rm f},t+1\right) = -g_i^+\left(\bm{x}_{\rm f},t\right) + 2\omega_i c_{\rm w}.
\label{eq:Zhang_stationary_walls}
\end{equation}
where $\bm{{\rm e}}_{i'} = -\bm{{\rm e}}_i$ and $g_i^+$ the post-collision distribution function.
\section{Results and discussions}
\label{sec:results_discussions}
\subsection{Effect of the flow on the released solute distribution}
\label{subsec:flow_regimes}
\begin{figure}[H]
\centering
\includegraphics[width = 0.7\textwidth, trim={1cm 3.58cm 1cm 2.4cm}, clip]{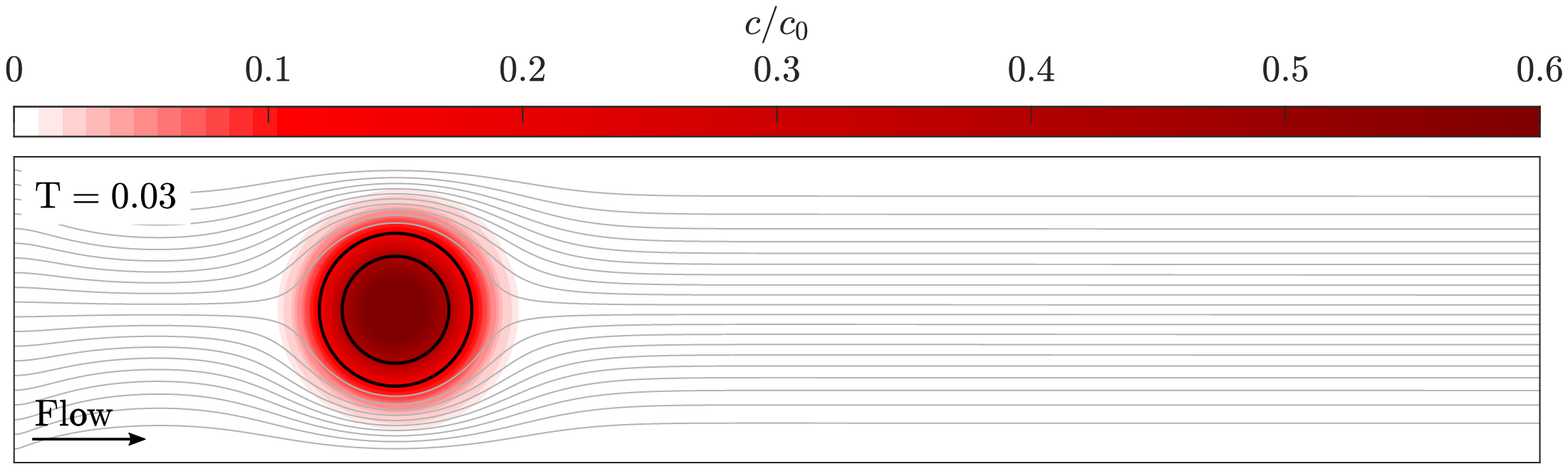}
\includegraphics[width = 0.7\textwidth, trim={1cm 4.4cm 1cm 4.4cm}, clip]{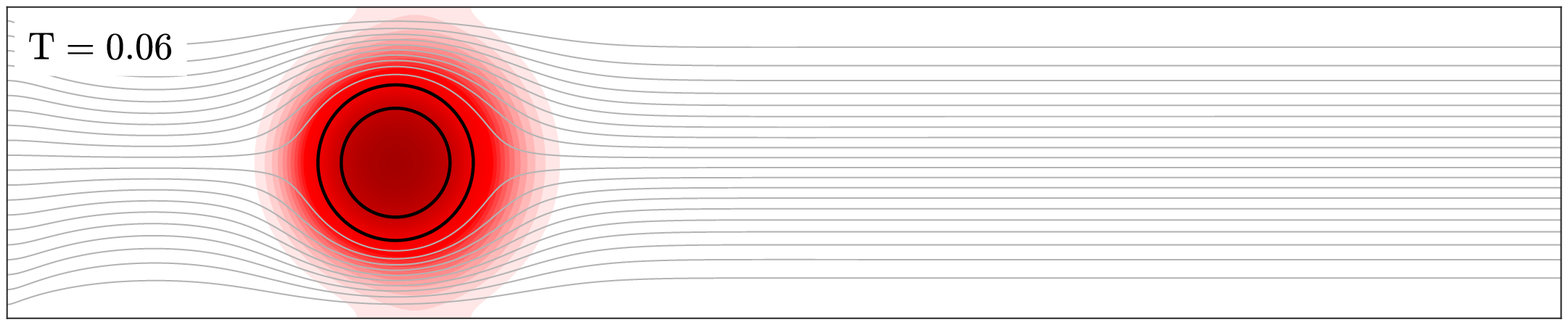}
\includegraphics[width = 0.7\textwidth, trim={1cm 4.4cm 1cm 4.4cm}, clip]{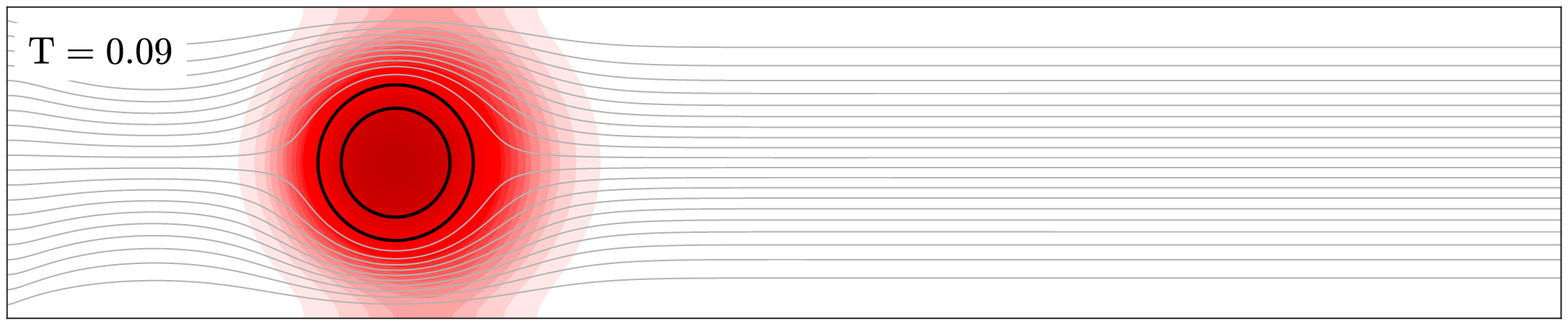}
\caption{Flow streamlines (grey solid lines) and solute concentration field (colormap) computed for ${\rm Re} = 1$, ${\rm B}=0.5$, $\rm Sc = 5$ and ${\rm P} = 3.70$. 
The core-shell cylinder is represented by the two black solid circles that stake out the inner and the outer interfaces of the shell.
Dark red corresponds to regions with maximal scaled solute concentration $c/c_0$, while white to absence of solute.
The flow is steady and laminar and it is from left to right.
Mass transfer is driven mainly by diffusion and takes place almost radially with respect to the core-shell cylinder axis.
The solute is barely advected by the flow; and thus, the concentration boundary layer thickness expands unsteadily until touching the channel walls.}
\label{fig:flow_mass_transfer_Re_1}
\end{figure}

\begin{figure}[H]
\centering
\includegraphics[width = 0.7\textwidth, trim={1cm 3.58cm 1cm 2.4cm}, clip]{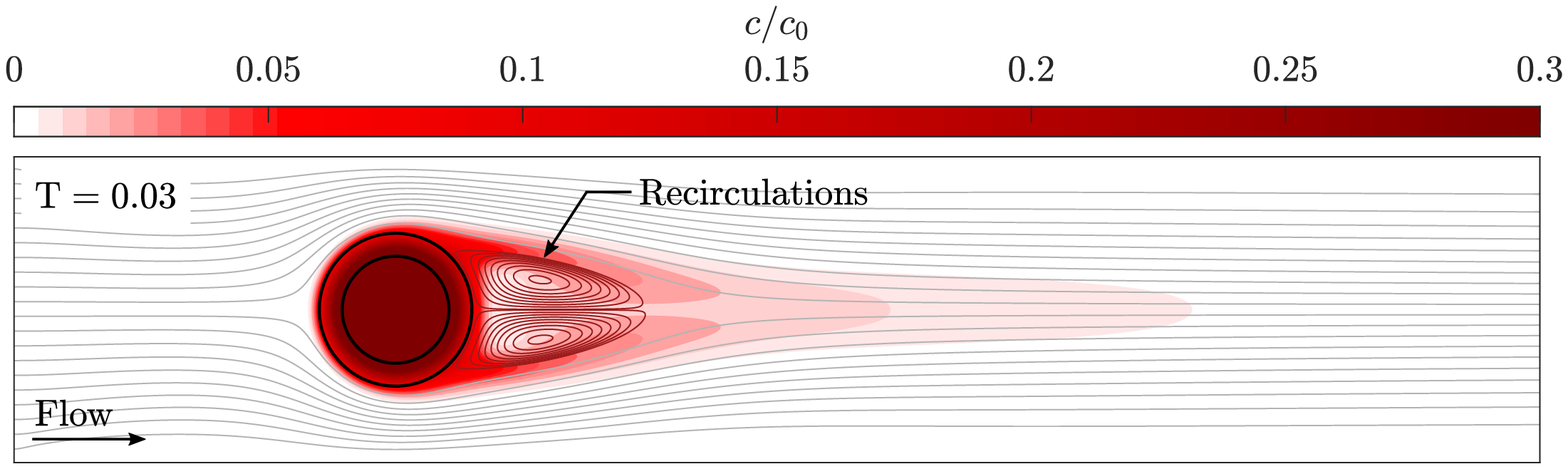}
\includegraphics[width = 0.7\textwidth, trim={1cm 4.4cm 1cm 4.4cm}, clip]{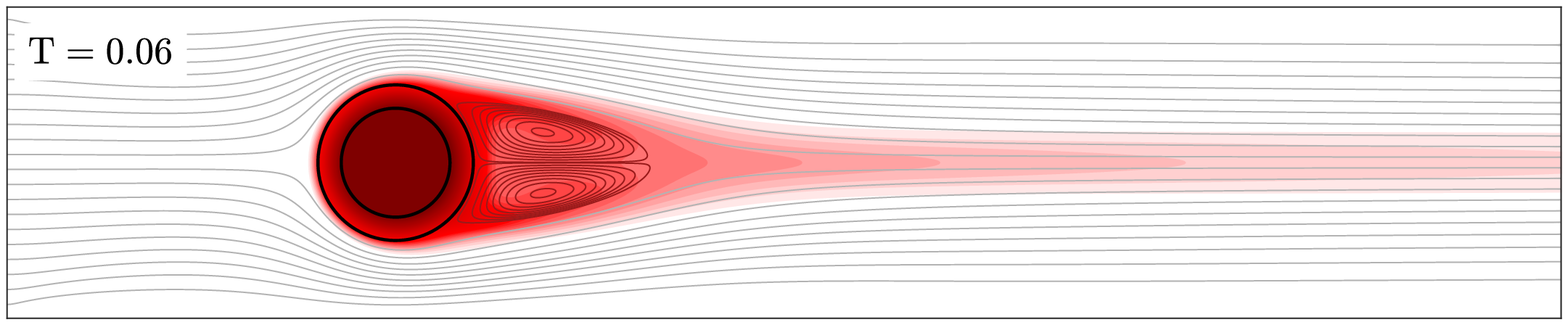}
\includegraphics[width = 0.7\textwidth, trim={1cm 4.4cm 1cm 4.4cm}, clip]{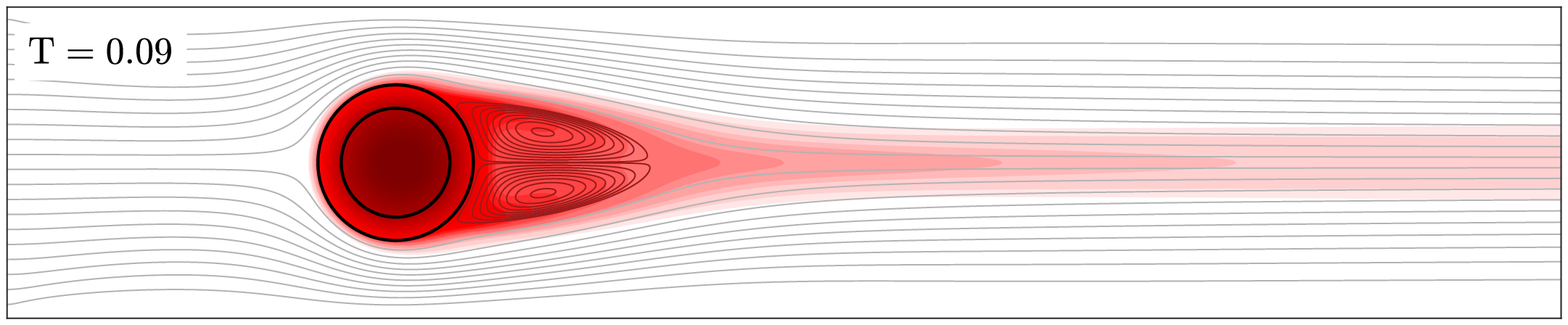}
\caption{Flow streamlines and solute concentration field computed at ${\rm Re} = 80$, ${\rm B}=0.5$, $\rm Sc = 5$ and ${\rm P} = 3.70$.
Steady recirculations emerge at the rear of the core-shell cylinder, where most of the released solute is trapped before it skips downstream by forming a long plume-shaped concentration front.
The isoconcentration contours around the reservoir are not steady, they move as the solute is released. The overall concentration front does not reach the channel walls due to increased efficacy of advection.}
\label{fig:flow_mass_transfer_Re_80}
\end{figure}

\begin{figure}[H]
\centering
\includegraphics[width = 0.7\textwidth, trim={1cm 3.58cm 1cm 2.4cm}, clip]{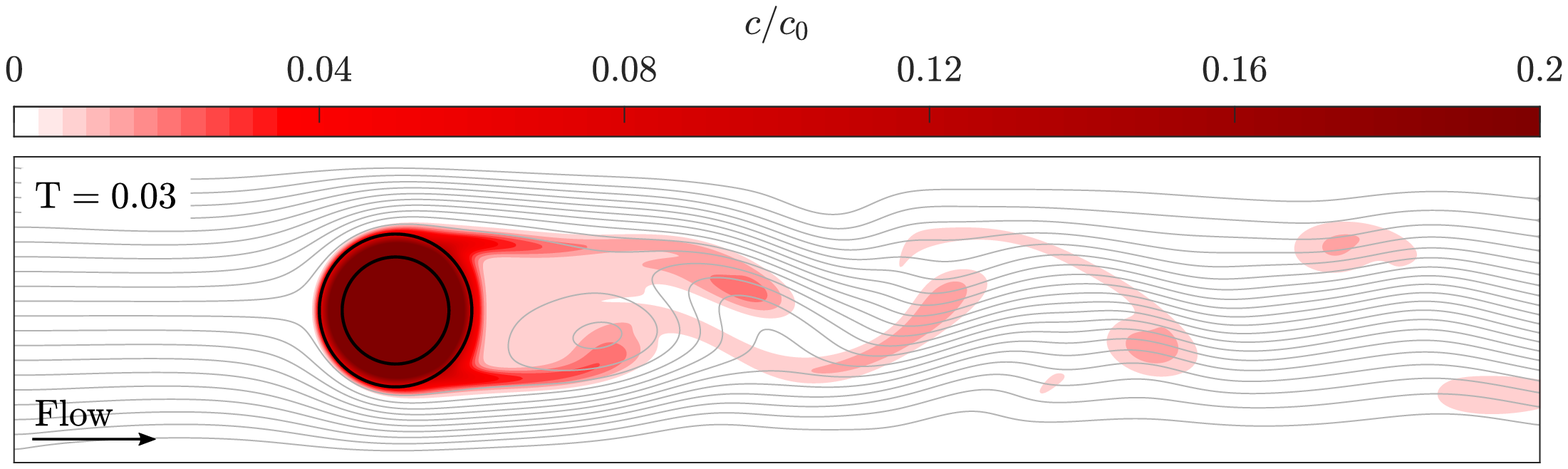}
\includegraphics[width = 0.7\textwidth, trim={1cm 4.4cm 1cm 4.4cm}, clip]{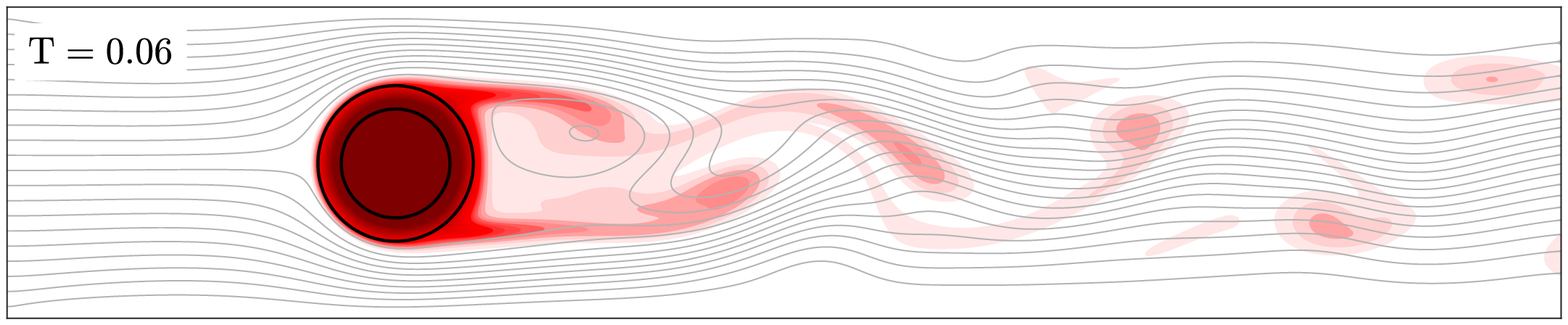}
\includegraphics[width = 0.7\textwidth, trim={1cm 4.4cm 1cm 4.4cm}, clip]{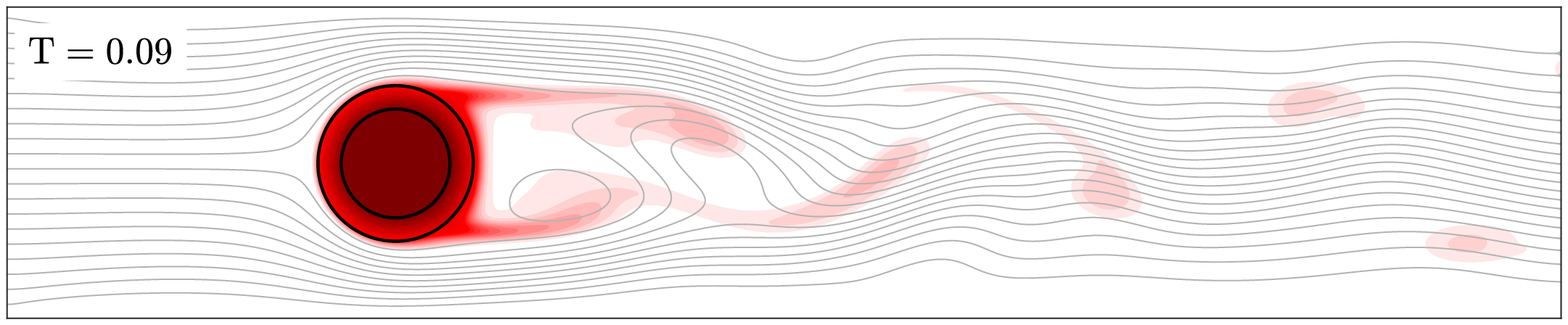}
\caption{Flow streamlines and solute concentration field computed at ${\rm Re} = 180$, ${\rm B}=0.5$, ${\rm P}= 3.70$ and $\rm Sc = 5$.
Flow is unsteady and exhibits a von K\'{a}rm\'{a}n vortex street that affects the spatial distribution and transport of the released solute.
Each vortex traps and accumulates shortly a small amount of solute, which decays later downstream.}
\label{fig:flow_mass_transfer_Re_180}
\end{figure}

Because the cylinder, in this study, is a reservoir whose load is not maintained constant in the course of time, the mass transfer needs to be solved outside as well as inside the reservoir.
Most classical studies have dealt with solving the solute mass transfer only outside the particles because the concentration is maintained constant inside as well as at the surface of the particles (see, \textit{e.g.}, \cite{Clift1978,Michaelides2006}).
Figures \ref{fig:flow_mass_transfer_Re_1}, \ref{fig:flow_mass_transfer_Re_80} and \ref{fig:flow_mass_transfer_Re_180} show the three typical fully developed flow patterns computed at $\mathrm{Re} = 1, 80$ and $180$, represented by the streamlines (the grey solid lines), and their induced solute spatial distribution. 
The two black solid circles represent the inner and the outer interfaces of the shell.
For these simulations, $\mathrm{Sc} = 5$ and $\mathrm{P} = 3.70$.
In each figure, the upper bond of the colorbar drops with increasing ${\rm Re}$.
The dark red color corresponds to regions with maximal solute concentration, while the white color to the absence of solute.
The transition of the flow from one regime to another is triggered by varying solely the Reynolds number ${\rm Re}$, by tuning the speed of the flow $U_{\infty}$, while holding all the other parameters constant.
At lower Reynolds numbers (${\rm Re} < 21$), the flow is laminar, steady and unseparated. 
Its streamlines exhibit almost a fore-aft symmetry, and are parallel to the channel walls far from the cylinder location away downstream.
At intermediate values ($21 < {\rm Re} < 128$), the hydrodynamic boundary layer detaches from the cylinder surface to form closed steady wake recirculations at the rear of the reservoir.
Beyond a critical value (${\rm Re} > 128$), the flow becomes unsteady and develops the famous periodically oscillating flow patterns, known as von K\'{a}rm\'{a}n vortex street. 
The transition depends on the blockage degree \cite{Shair1963}.
The actual measured tresholds, ${\rm Re}=21$ and ${\rm Re}=128$, are close to the values reported in Refs.~\cite{Chen1995,Sahin2004} for a confined cylinder at ${\rm B} = 0.5$.
The blockage alters the hydrodynamic boundary layer and the streamlines configuration at the vicinity of the reservoir.

For $\mathrm{Re} < 21$, the concentration in the core decays noticeably over time as the reservoir is leaking its initial loaded solute, as shown in Fig.~\ref{fig:flow_mass_transfer_Re_1} at $\mathrm{Re} = 1$.
At lower Reynolds numbers, the solute transport is dominated by diffusion even in the surrounding flowing fluid.
The solute diffuses rapidly around the cylinder, and it is barely advected downstream.
The concentration boundary layer is unsteady, and it continues to expand until it touches the channel walls.
For $21 < {\rm Re} < 128$, the concentration boundary layer is thin at the front of the cylinder, and develops a long plume downstream.
These emerging features are shown in Fig.~\ref{fig:flow_mass_transfer_Re_80} for the case of ${\rm Re} = 80$.
In this situation, the mass transport in the fluid is dominated by advection that transports the released solute downstream until the channel outlet.
The released solute at the front of the reservoir is instantaneously washed away by the newly upcoming flow, which leads to the emergence of a steady thin concentration boundary layer there and up to the flow separation points. 
Most of the solute is trapped in the wake vortices, where it accumulates for a while before it skips out by diffusion. 
Here, the isoconcentration contours around the reservoir are unsteady because neither the concentration nor the mass flux are maintained at constant values on the reservoir surface.
They keep evolving until the reservoir loses completely its load.
For ${\rm Re} > 128$, the von K\'{a}rm\'{a}n vortex street affects the spatial distribution of the released solute, as shown in Fig.~\ref{fig:flow_mass_transfer_Re_180} for $\mathrm{Re} = 180$.
The concentration boundary layer is now even thinner at the front of the cylinder because the flow velocity is larger, and so does advection. 
The solute never reaches the channel walls when $\mathrm{Re} > 21$; and thus, the concentration boundary layer thickness is not affected by the presence of the walls at larger $\mathrm{Re}$, for which the mass transfer correlation is extracted later on. 
The reported observations of the effect of steady and unsteady flows on the spatial distribution of the released solute has not been reported elsewhere.
The previously reported results computed in the limit of Stokes flows \cite{Clift1978}, and for unsteady flows \cite{Karniadakis1988,Bouhairi2007,Haeri2013} around a circular cylinder are obtained for constant surface concentration boundary condition (constant temperature in the previous studies).
This simplifies the calculations by limiting the computation only to the outer domain.

\begin{figure}[H]
\centering
\includegraphics[width = 0.7\textwidth, trim={1cm 3.58cm 1cm 2.4cm}, clip]{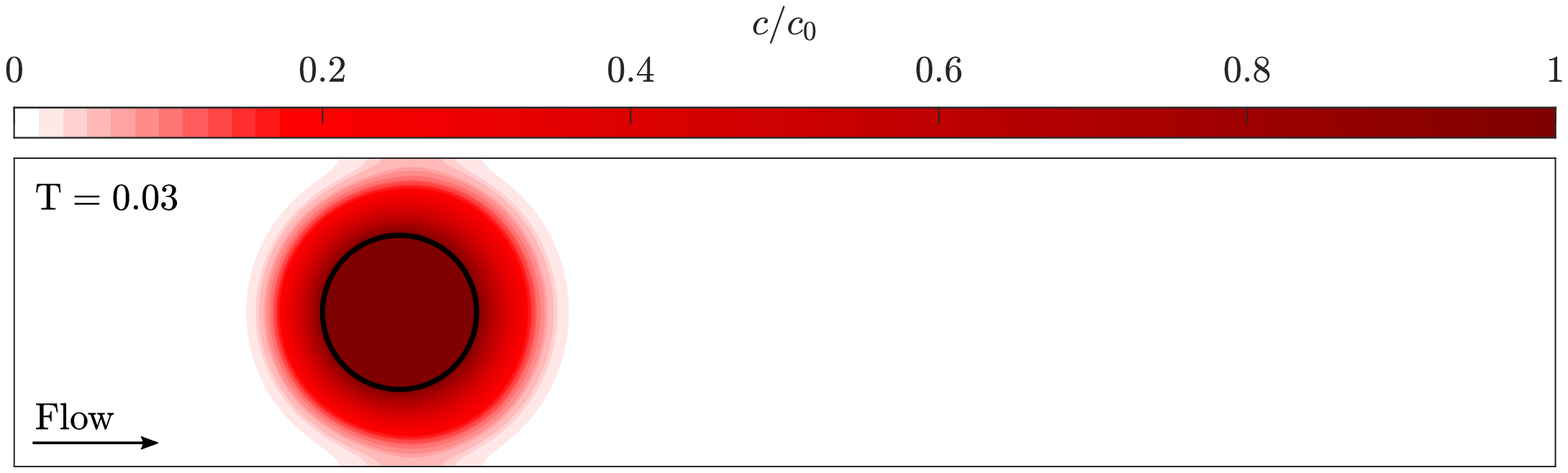}
\includegraphics[width = 0.7\textwidth, trim={1cm 4.4cm 1cm 4.4cm}, clip]{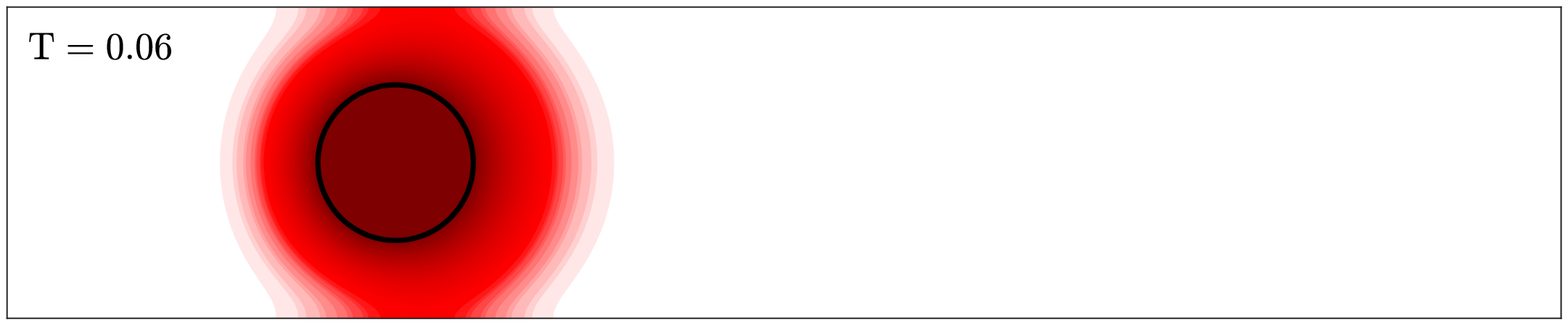}
\includegraphics[width = 0.7\textwidth, trim={1cm 4.4cm 1cm 4.4cm}, clip]{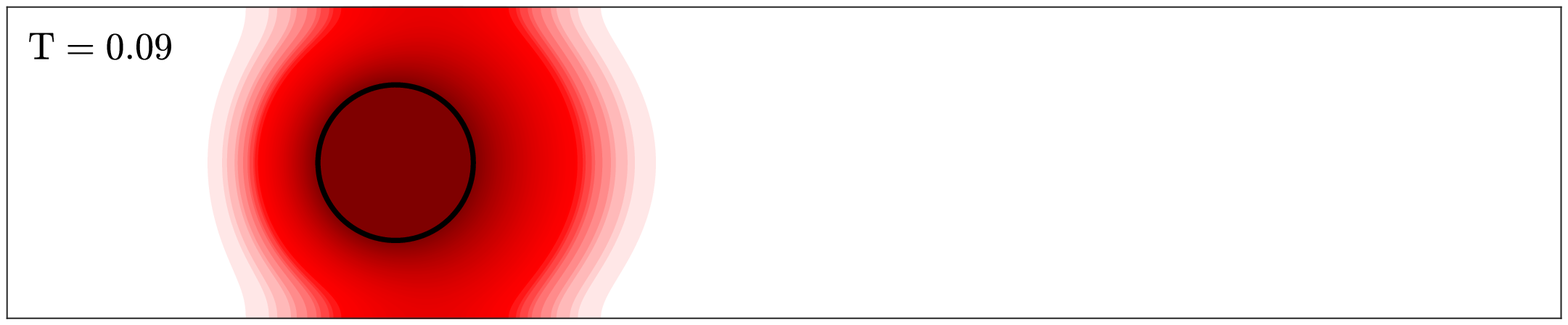}
\caption{Solute concentration field around a cylinder in crossflow whose surface is maintained at uniform and constant concentration $c_{\rm w}$, in absence of any coating shell. ${\rm Re} = 1$, ${\rm B}=0.5$ and $\rm Sc = 5$. The background applied flow is the same as in Fig.~\ref{fig:flow_mass_transfer_Re_1}.
The solute mass transfer is unsteady, and mainly dominated by diffusion.
The diffusion front spreads faster compared to the situation of the cylindrical core-shell reservoir reported in Fig.~\ref{fig:flow_mass_transfer_Re_1}.
The maximum of $c/c_0$ remains at $1$ all the time.}
\label{fig:concentration_field_Re_1}
\end{figure}

\begin{figure}[H]
\centering
\includegraphics[width = 0.7\textwidth, trim={1cm 3.58cm 1cm 2.4cm}, clip]{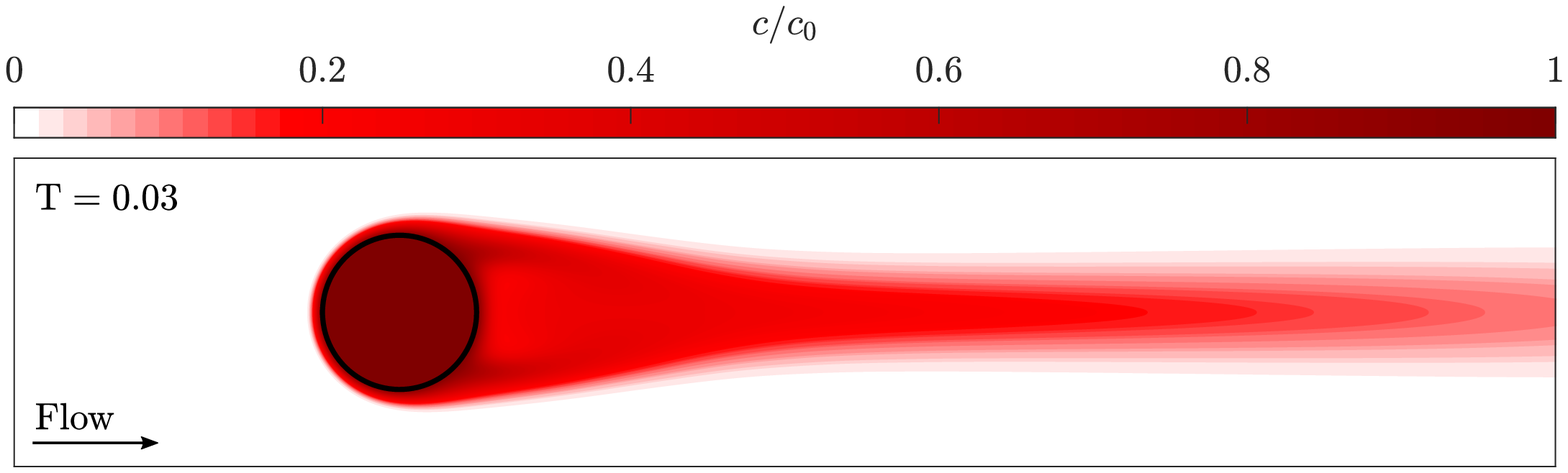}
\includegraphics[width = 0.7\textwidth, trim={1cm 4.4cm 1cm 4.4cm}, clip]{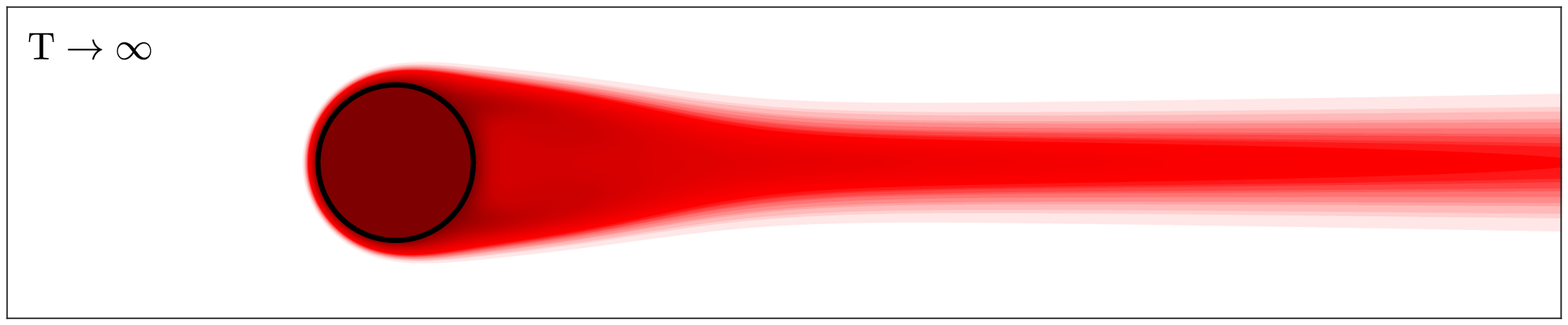}
\caption{
Solute concentration field around a cylinder whose surface is maintained at uniform and constant concentration $c_{\rm w}$, in crossflow and in absence of any coating shell. ${\rm Re} = 80$, ${\rm B}=0.5$ and $\rm Sc = 5$. The background applied flow is the same as in Fig.~\ref{fig:flow_mass_transfer_Re_80}.
Both the flow and the mass transfer reach a steady regime.
The concentration boundary layer adopts a steady thickness and the plume at the rear of the cylinder is also steady, and its concentration is higher compared to the situation of the core-shell reservoir reported in Fig.~\ref{fig:flow_mass_transfer_Re_80}.
}
\label{fig:concentration_field_Re_80}
\end{figure}

\begin{figure}[H]
\centering
\includegraphics[width = 0.7\textwidth, trim={1cm 3.58cm 1cm 2.4cm}, clip]{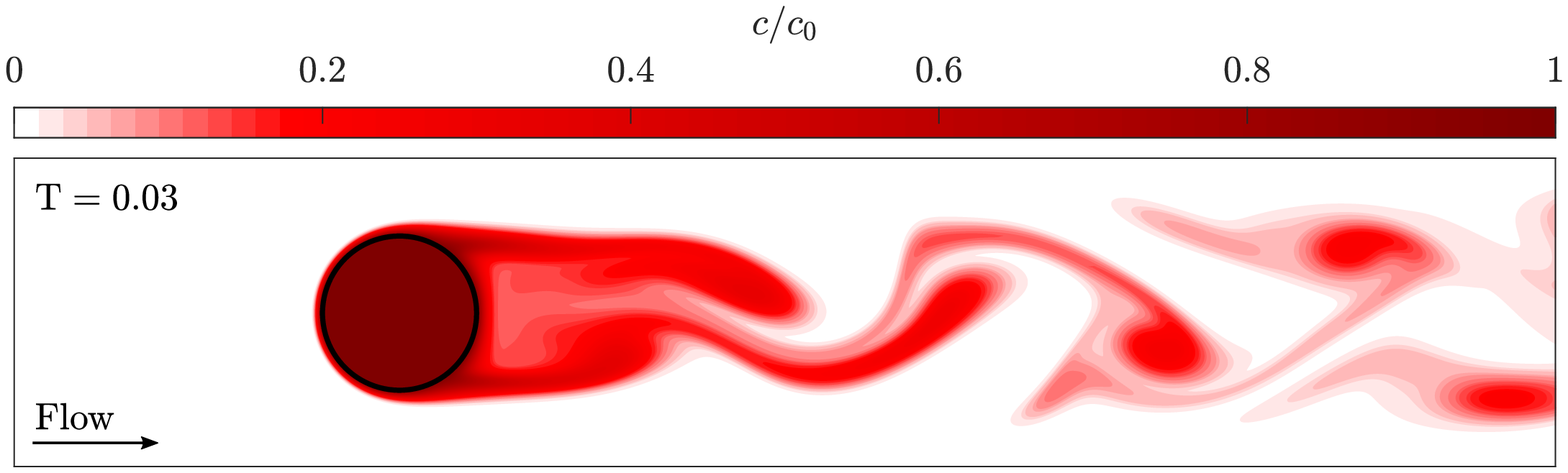}
\includegraphics[width = 0.7\textwidth, trim={1cm 4.4cm 1cm 4.4cm}, clip]{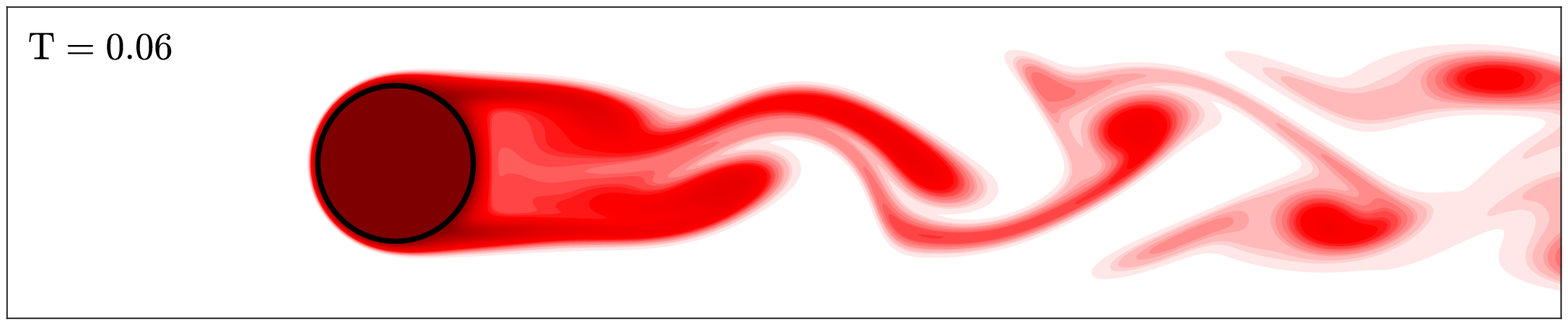}
\includegraphics[width = 0.7\textwidth, trim={1cm 4.4cm 1cm 4.4cm}, clip]{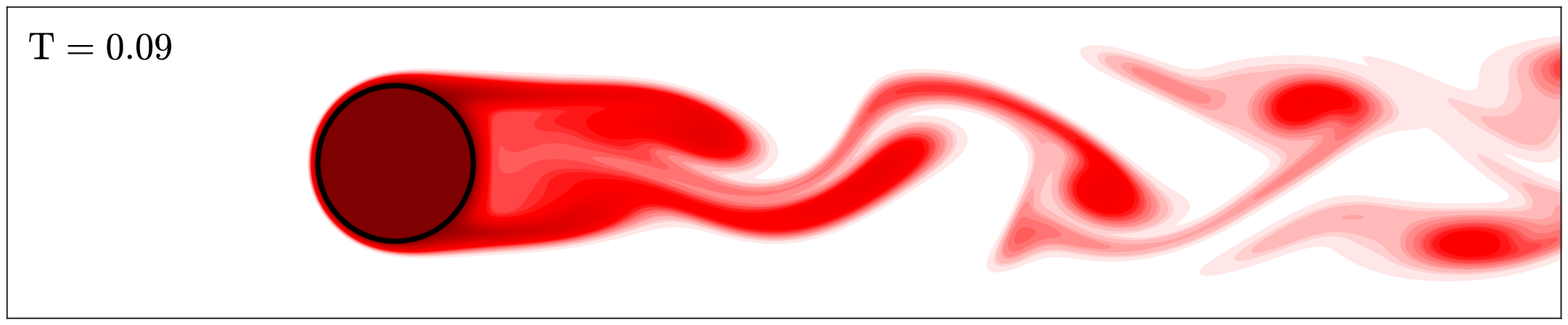}
\caption{
Solute concentration field around a cylinder whose surface is maintained at uniform and constant concentration $c_{\rm w}$, in crossflow and in absence of any coating shell. ${\rm Re} = 180$, ${\rm B}=0.5$ and $\rm Sc = 5$. The background applied flow is the same as in Fig.~\ref{fig:flow_mass_transfer_Re_180}.
Because the concentration at the surface of the cylinder is kept constant, the resulting solute concentration distribution is higher (with darker red colored regions) and undergoes sustained oscillation, compared to the situation of a cylindrical core-shell reservoir reported in Fig.~\ref{fig:flow_mass_transfer_Re_180}.
}
\label{fig:concentration_field_Re_180}
\end{figure}

For comparison purpose, simulations with constant concentration $c_{\rm w}=1$ at the surface of a cylinder - without a shell - are also conducted in this study at the same Schmidt number $\mathrm{Sc = 5}$ and Reynolds numbers $\mathrm{Re} = 1$, $80$ and $180$.
Snapshots taken at similar dimensionless times, as in Figures \ref{fig:flow_mass_transfer_Re_1}, \ref{fig:flow_mass_transfer_Re_80} and \ref{fig:flow_mass_transfer_Re_180}, are shown in Figs.~\ref{fig:concentration_field_Re_1}, \ref{fig:concentration_field_Re_80} and \ref{fig:concentration_field_Re_180}.
They allow to appreciate the differences between having the classically imposed constant concentration boundary condition at the cylinder surface, and having the continuous boundary conditions involved by the reservoir (Eqs.~\ref{eq:bc_cs} and \ref{eq:bc_sf}).
\newpage
\newpage
Mass transfer is clearly dominated by diffusion again at $\mathrm{Re} = 1$, while it is dominated by advection at $\mathrm{Re} = 80$ and $180$. 
However, Figs.~\ref{fig:concentration_field_Re_1}, \ref{fig:concentration_field_Re_80} and \ref{fig:concentration_field_Re_180} exhibit differences.
For example, the solute concentration in the vicinity of the cylinder and downstream is higher for the isoconcentration boundary condition.
The maximal concentration in all the snapshots is the same because the concentration inside the cylinder does not decrease over time.
It is maintained constant at $c_0=c_{\rm w}=1$.
The system evolves to a steady state at ${\rm Re}=80$ (Fig.~\ref{fig:concentration_field_Re_80}), in contrast to the case of the reservoir (Fig.~\ref{fig:flow_mass_transfer_Re_80}).
For ${\rm Re}=180$, the released solute develops sustained periodical moving patterns downstream (Fig.~\ref{fig:concentration_field_Re_180}), while the overall concentration decays over time in Fig.~\ref{fig:flow_mass_transfer_Re_180}.
\newpage
\newpage
\subsection{Effect of the flow on the local mass transfer quantities}
\label{subsec:local_quantities}
\begin{figure}[H]
\subfloat[\label{subfig:local_concentration_Re_1}]{\includegraphics[width=0.32\textwidth]{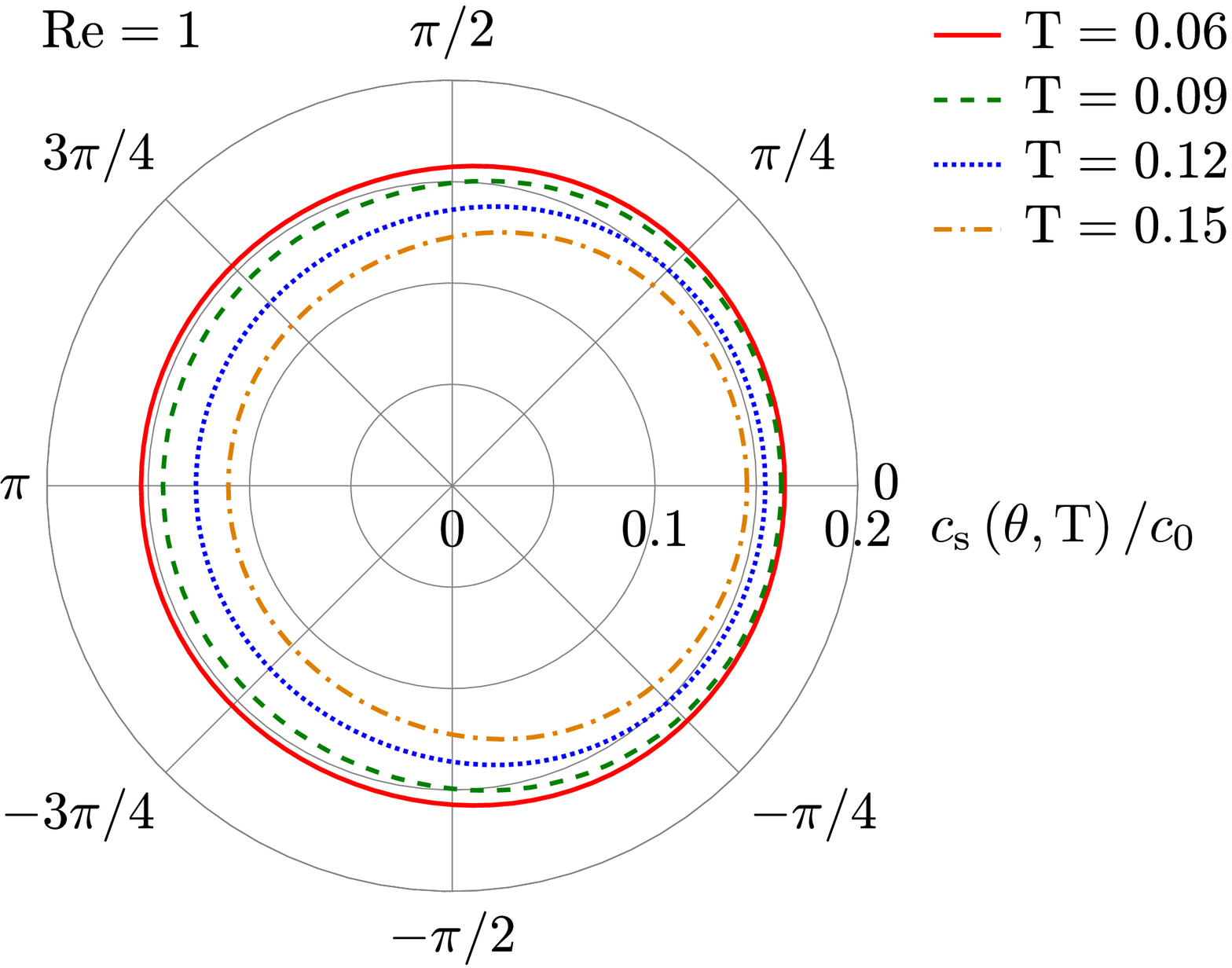}}
\hfill
\subfloat[\label{subfig:local_concentration_Re_80}]{\includegraphics[width=0.32\textwidth]{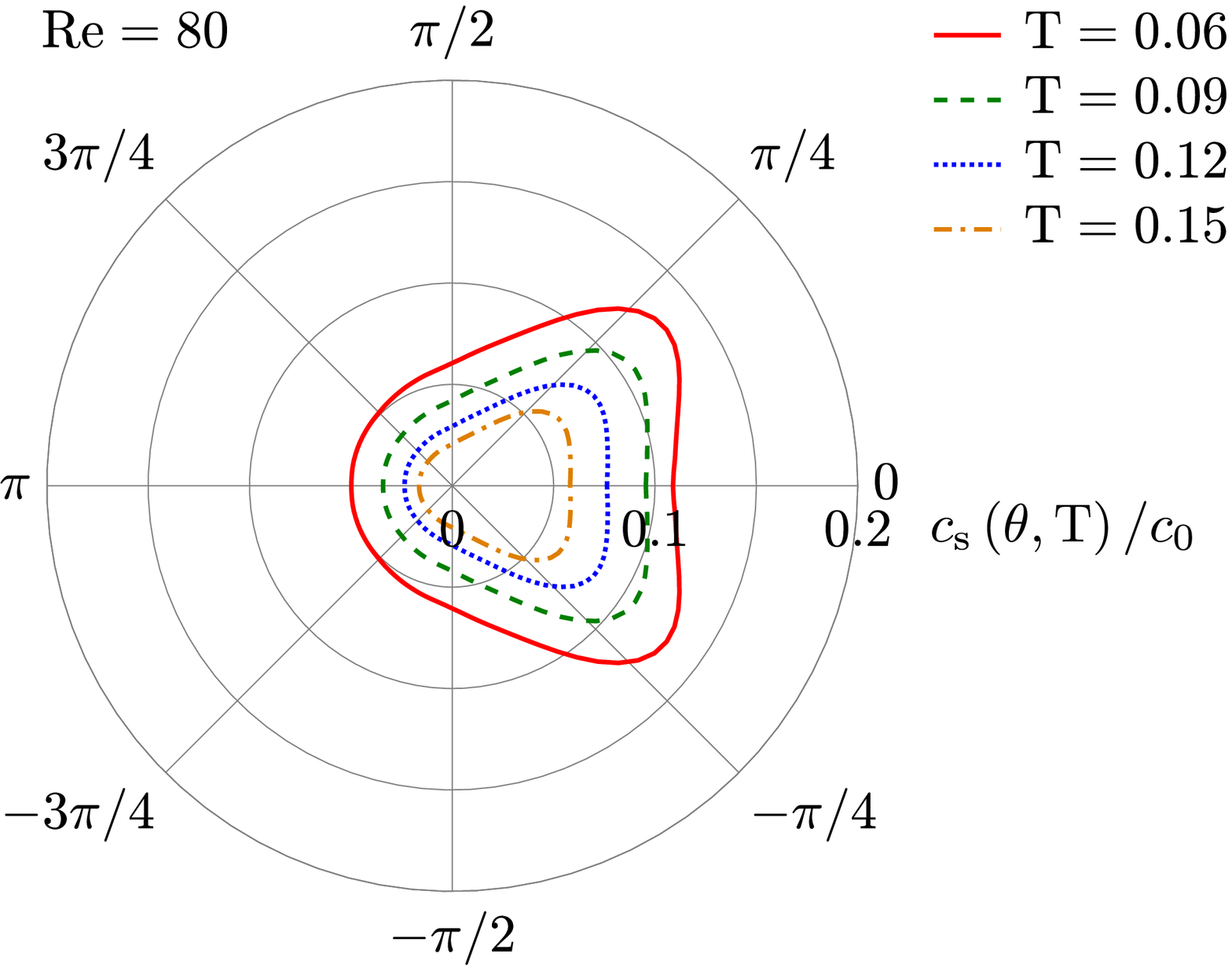}}
\hfill
\subfloat[\label{subfig:local_concentration_Re_180}]{\includegraphics[width=0.32\textwidth]{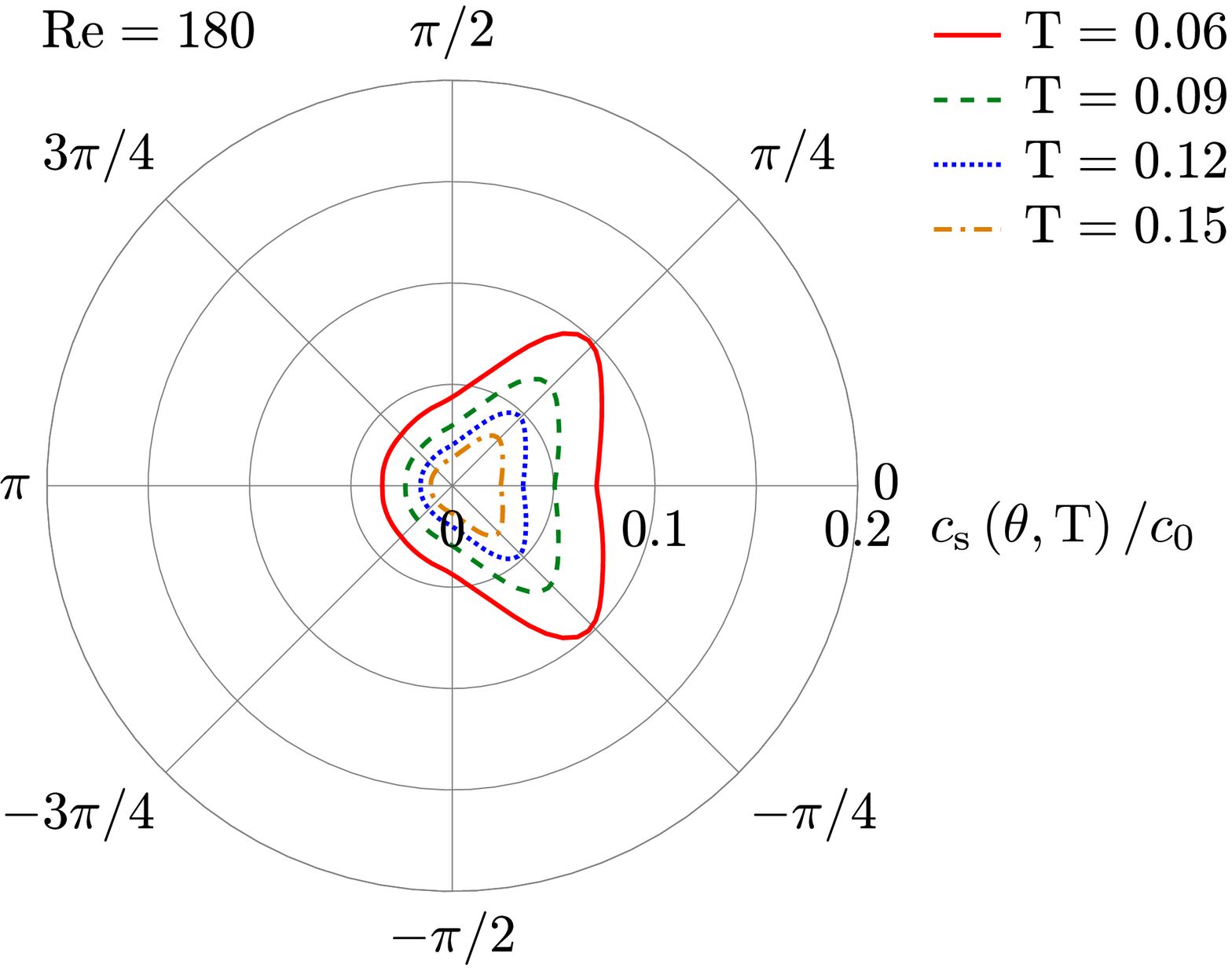}}     
\caption{Solute concentration $c_{\rm s}(\theta,{\rm T})$ at the surface of the cylindrical core-shell reservoir under crossflow at various Reynolds numbers ${\rm Re}$, while holding the same Schmidt number ${\rm Sc} = 5$ and the same dimensionless shell solute permeability ${\rm P} = 3.70$. Distortion of the concentration uniformity is largely amplified by increasing the flow speed.}
\label{fig:local_concentration}
\end{figure}

\begin{figure}[H]
\subfloat[\label{subfig:local_flux_Re_1}]{\includegraphics[width=0.32\textwidth]{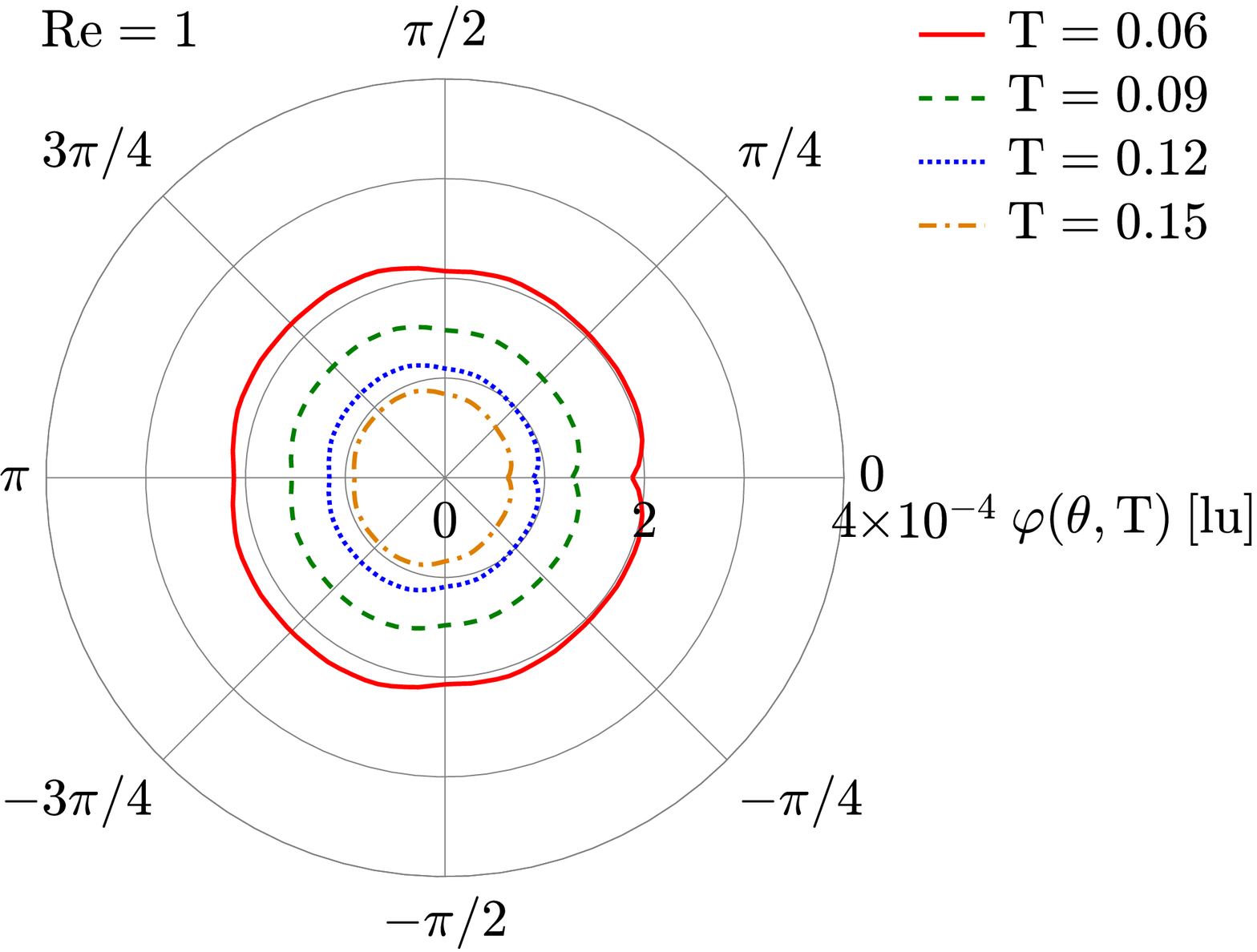}}
\hfill
\subfloat[\label{subfig:local_flux_Re_80}]{\includegraphics[width=0.32\textwidth]{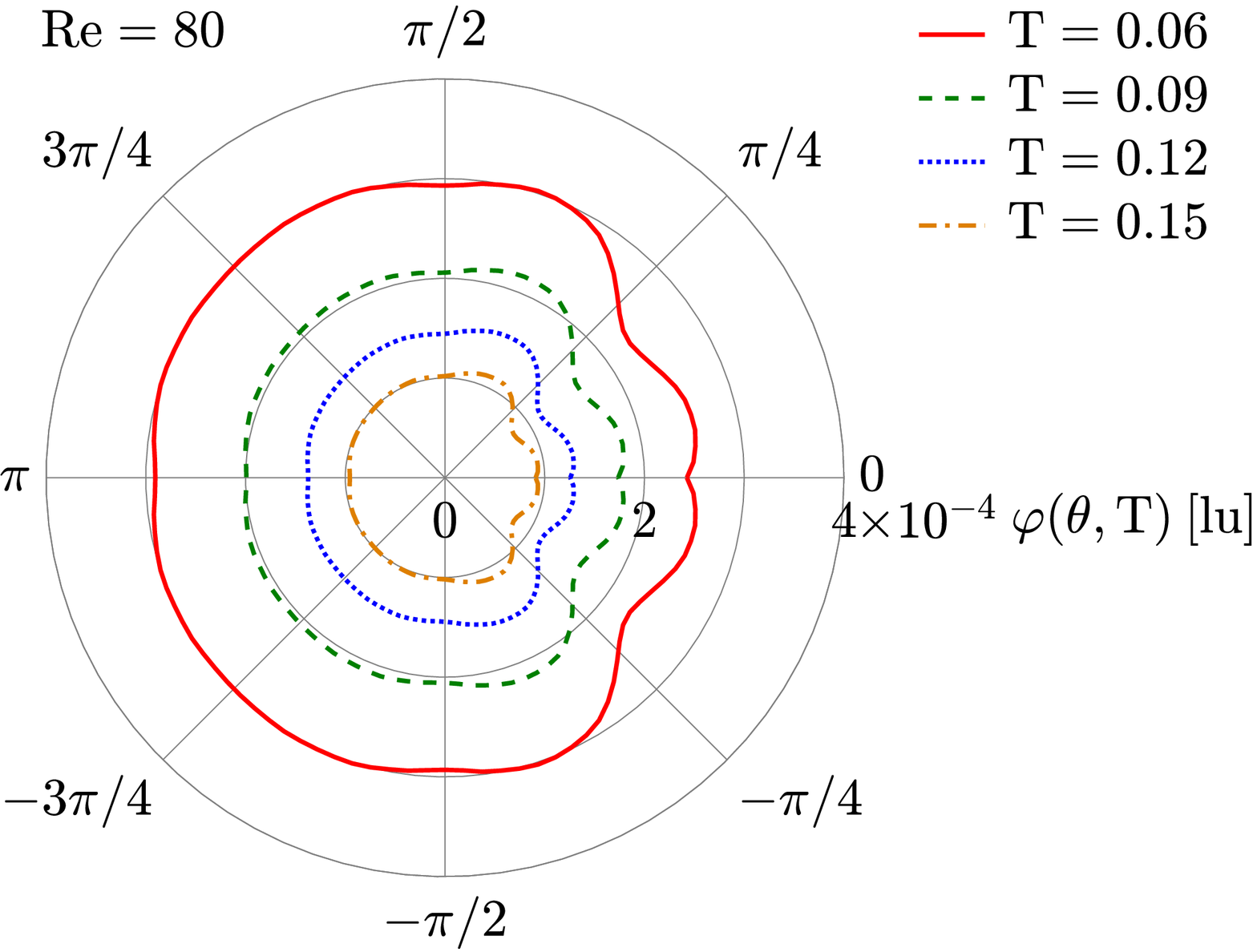}}
\hfill
\subfloat[\label{subfig:local_flux_Re_180}]{\includegraphics[width=0.32\textwidth]{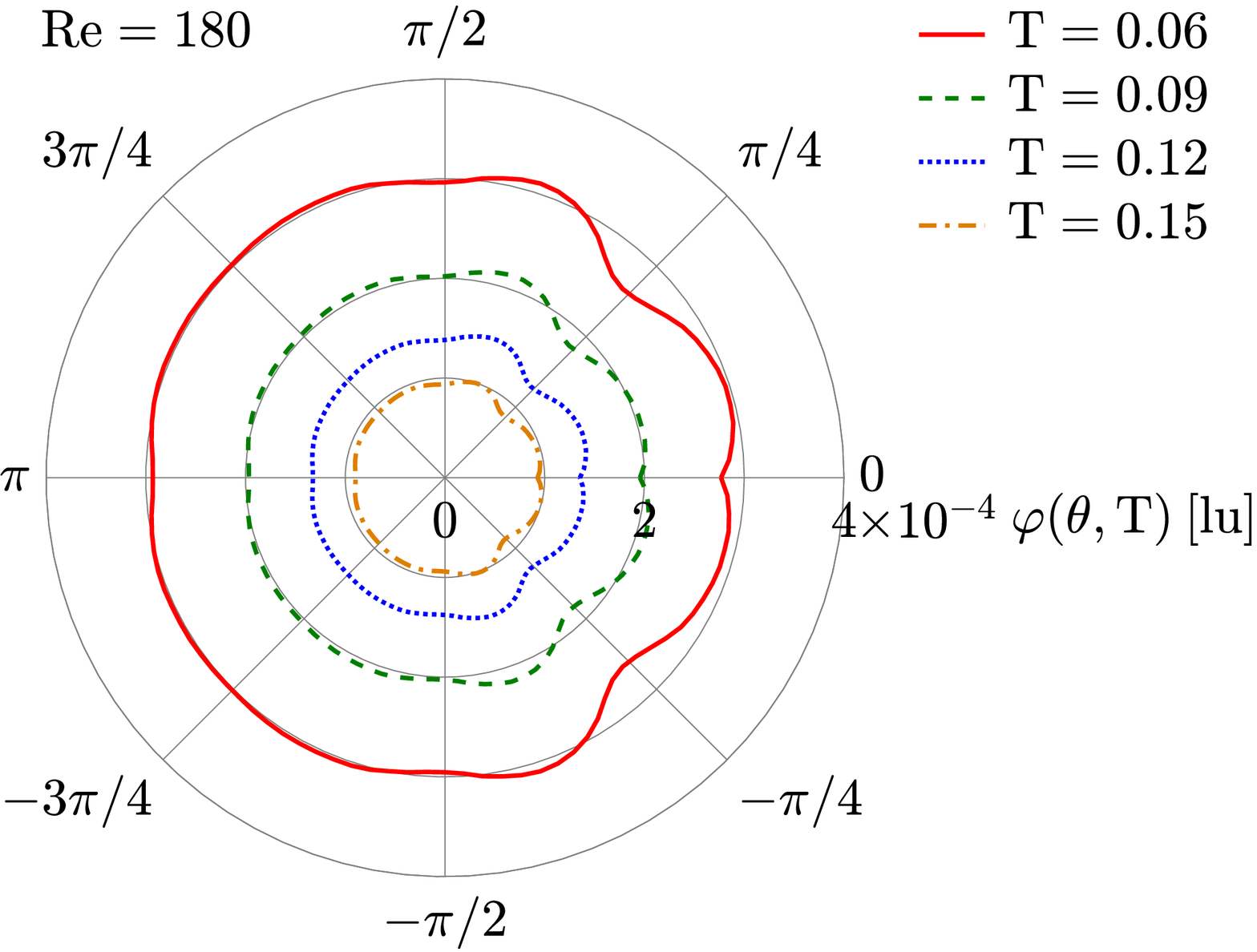}}     
\caption{(a,b,c) The local mass flux at the surface of the reservoir $\varphi(\theta,{\rm T})$ at different Reynolds numbers, while holding the same Schmidt number ${\rm Sc} = 5$ and the same dimensionless shell solute permeability ${\rm P} = 3.70$. $\varphi(\theta,{\rm T})$ is amplified by increasing ${\rm Re}$. It is almost uniform, with a localized drop at the separation points when ${\rm Re}$ is large. $\varphi(\theta,{\rm T})$ is expressed in lattice units.}
\label{fig:local_flux}
\end{figure}

\begin{figure}[H]
\subfloat[\label{subfig:local_Sherwood_Re_1}]{\includegraphics[width=0.32\textwidth]{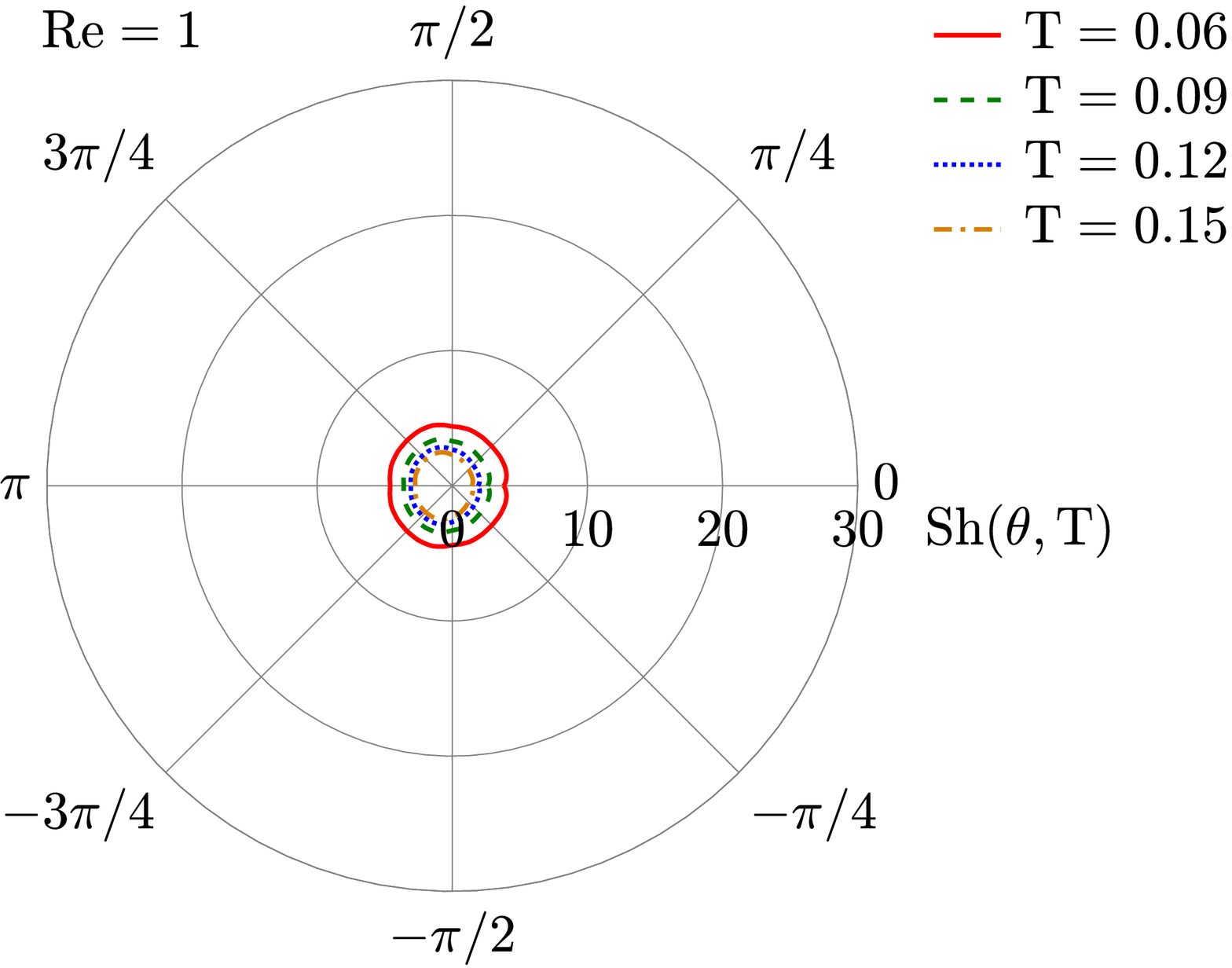}}
\hfill
\subfloat[\label{subfig:local_Sherwood_Re_80}]{\includegraphics[width=0.32\textwidth]{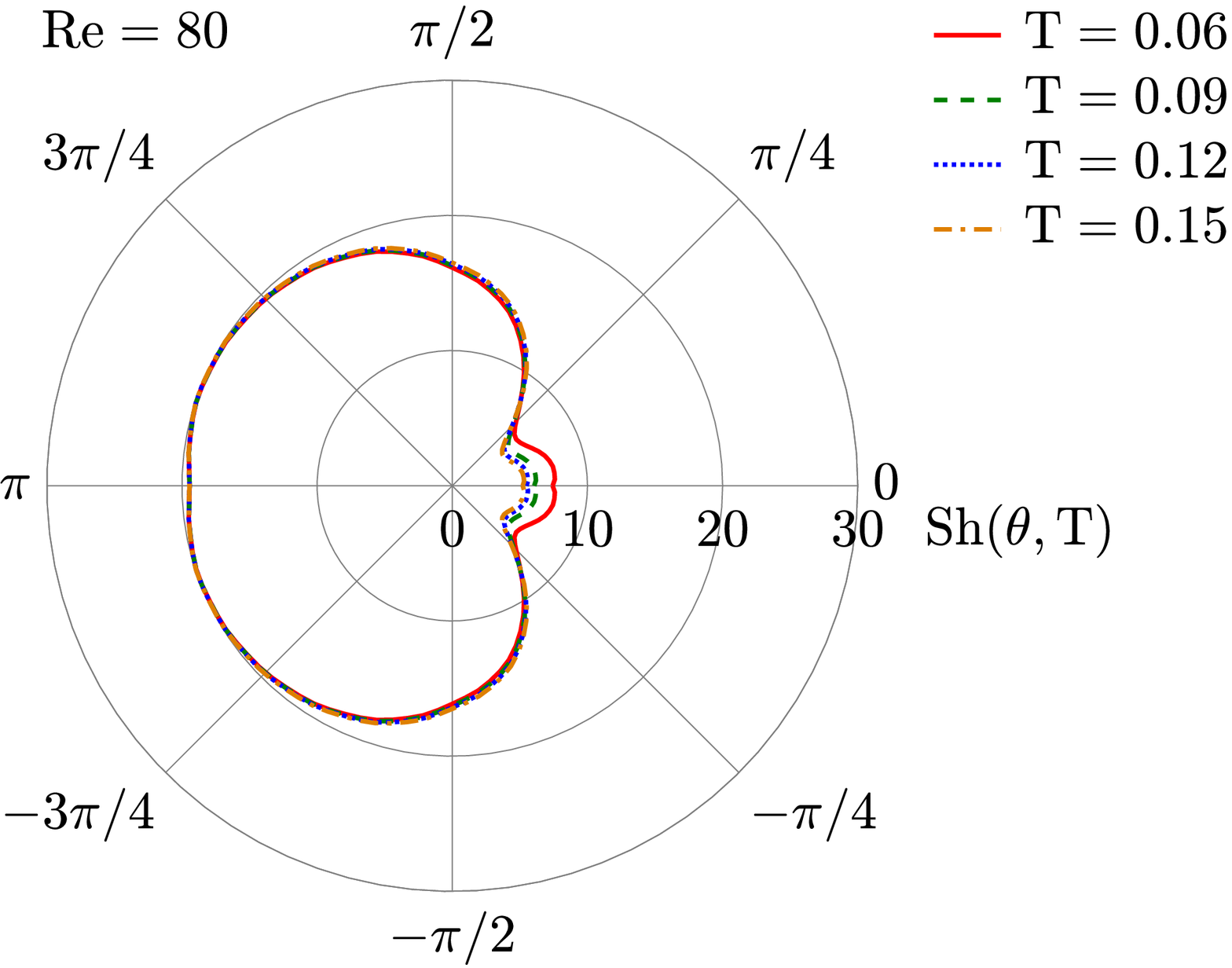}}
\hfill
\subfloat[\label{subfig:local_Sherwood_Re_180}]{\includegraphics[width=0.32\textwidth]{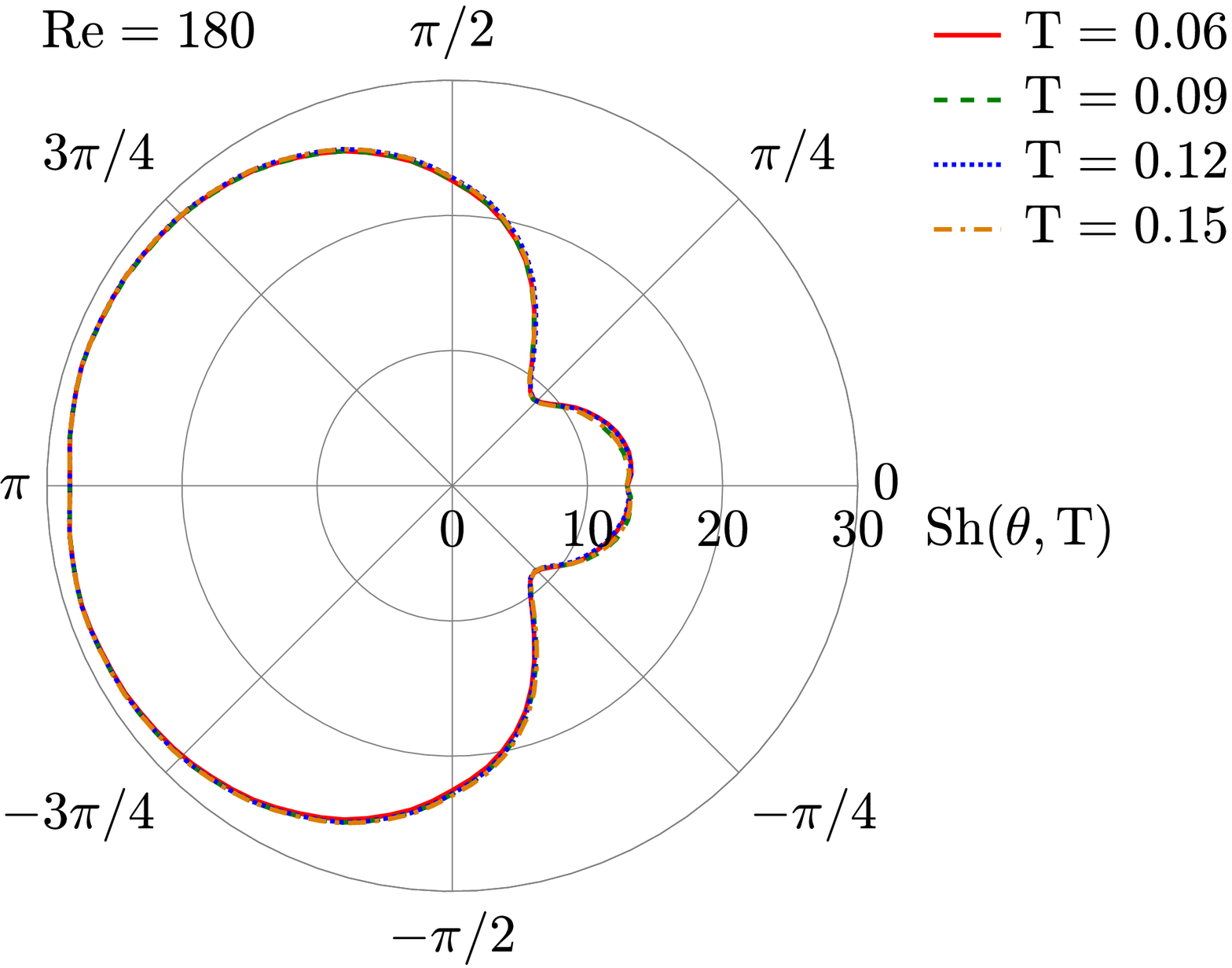}}     
\caption{Instantaneous local Sherwood number ${\rm Sh(\theta,{\rm T})}$ at the surface of a cylindrical core-shell reservoir in crossflow at various Reynolds numbers. Both the Schmidt number and the scaled shell solute permeability are hold constant, ${\rm Sc} = 5$ and ${\rm P} = 3.70$. At large Reynolds numbers, ${\rm Sh(\theta,{\rm T})}$ is steady.
Similar qualitative, but not quantitative, behavior is observed for cylinders without a shell and with constant surface concentration \cite{Krall1973,Karniadakis1988}).}
\label{fig:local_Sherwood}
\end{figure}

The essential of mass transfer occurs at the surface of the core-shell reservoir, which is in contact with the ambient fluid, and where the local concentration $c_{\rm s}(\theta,{\rm T})$, the local mass flux $\varphi(\theta,{\rm T})$ and the resulting local Sherwood number $\rm{Sh}(\theta,{\rm T})$ are found to evolve differently depending on the flow regime.
Figures \ref{fig:local_concentration}, \ref{fig:local_flux} and \ref{fig:local_Sherwood} report in polar coordinates the circumferential mass transfer quantities, measured at Reynolds numbers ${\rm Re} = 1$, $80$ and $180$.
Only data for ${\rm Sc} = 5$ are shown, while similar behavior is observed for other values within the range $1 \leq \mathrm{Sc} \leq 25$ explored so far in this study.
The flow is from left to right, with $\theta=\pi$ located on the front of the reservoir upstream, and $\theta=0$ at its rear downstream.
In all cases, the overall concentration in the reservoir tends to vanish at long-term as the encapsulated solute is completely released out (equilibrium state).
At any given time, the local concentration is lower at high Reynolds numbers as the solute release is enhanced by the flow.
This corresponds to speeding up the exhaustion of the reservoir, from its initially loaded cargo, by increasing the flow velocity $U_\infty$.
This affects also how the solute is distributed at the surface of the reservoir.
The concentration at the surface of the reservoir $c_{\rm s}(\theta,{\rm T})$ shows only slight variations in space, at each time ${\rm T}$, when ${\rm Re}=1$.
It depends weakly on $\theta$.
However, at $\rm{Re} = 80$ and $\rm{Re} = 180$, $c_{\rm s}(\theta,{\rm T})$ depends strongly on $\theta$.
It varies nonmonotonically, from the front to the rear of the reservoir, while adopting two maximum values around $\theta = \pm\frac{\pi}{4}$ that corresponds to the separation points of the concentration boundary layer.
A common observed feature, at all Reynolds numbers, is that $c_{\rm s}(\theta,{\rm T})$ is always perfectly uniform at the front of the reservoir at the vicinity of $\theta = \pi$.
In this section of the reservoir surface, the thickness of the concentration boundary layer is maintained uniform as long as the flow is laminar, with streamlines locally parallel to the reservoir surface, and resulting in a high local pressure that pushes the solute against the reservoir surface.
The reported nonuniform and unsteady evolution in time of $c_{\rm s}(\theta,{\rm T})$ is a characteristic feature of having a reservoir, with an initial nonsustained solute load (Eq.~(\ref{eq:ic})) and the continuity of mass transfer boundary conditions at its surface (Eqs.~\ref{eq:bc_cs} and \ref{eq:bc_sf}).
Nonuniform, but steady, character of $c_{\rm s}(\theta,{\rm T})$ can be expected for cylinders without a shell when their surface is hold at constant mass flux.
Unfortunately these quantities are not usually reported in literature, instead only the local Sherwood number is reported (the Nusselt number in thermal flows Refs.~\cite{Krall1973,Karniadakis1988,Haeri2013}).
In this study, however, the local mass flux $\varphi(\theta,{\rm T})$, obtained by Eq.~(\ref{eq:local_flux}), is also nonuniform and unsteady.
Figure~\ref{fig:local_flux} shows how $\varphi(\theta,{\rm T})$ is uniform only at the front of the reservoir, for all the three flow regimes, and drops steeply down to a minimum around $\theta = \pm\frac{\pi}{4}$ (the separation points) at larger ${\rm Re}$.
$\varphi(\theta,{\rm T})$ gets larger values where the flow can advect the released solute downstream easily, and can maintain a steeper concentration gradient. 

In the present case study, both $c_{\rm s}(\theta,{\rm T})$ and $\varphi(\theta,{\rm T})$ are not steady and not uniform, in contrast to the previously studied cases in literature, where only one of them is set constant, as a control parameter, and the other evolves accordingly into a steady nonuniform distribution.
Here, both $c_{\rm s}(\theta,{\rm T})$ and $\varphi(\theta,{\rm T})$ are observable quantities and continue to evolve in time unsteadily until complete exhaustion of the reservoir from its initial solute load.
Figures~\ref{fig:local_concentration} and \ref{fig:local_flux} show how mass transfer tends to equilibrium without adopting a steady regime.

$c_{\rm s}(\theta,{\rm T})$ and $\varphi(\theta,{\rm T})$ are used in Eq.~(\ref{eq:local_sh}) to determine the local Sherwood number $\rm{Sh}(\theta,{\rm T})$, which is reported in Fig.~\ref{fig:local_Sherwood}. 
It is lower, uniform and unsteady at $\rm{Re} = 1$.
For $\rm{Re} = 80$, $\rm{Sh}(\theta,{\rm T})$ is larger at the front of the reservoir where the advection is important, and lower at the rear where the solute transport is operated mostly by diffusion.
$\rm{Sh}(\theta,{\rm T})$  does not vary in time on the upstream front of the reservoir because the released solute is constantly washed away by the flow, and thus, the concentration boundary layer thickness is maintained constant in time. 
$\rm{Sh}(\theta,{\rm T})$ gets minima at the separation points due to the local switch of the mass transfer mechanism from advection to diffusion dominated regime. 
For $\rm{Re} = 180$, $\rm{Sh}(\theta,{\rm T})$ is further amplified because of the increase in the advection contribution, and the further reduction in the concentration boundary layer thickness. 
$\rm{Sh}(\theta,{\rm T})$ is steady everywhere, even beyond the separation point. 
Interestingly, the Sherwood number of the core-shell reservoir is steady despite the unsteadiness of both the local concentration and the local mass flux, whose instantaneous ratio is; however, a constant.
The resulting local Sherwood number shows similar qualitative behavior as observed for a cylinder without a shell and having either constant concentration or constant mass flux at its surface (see, \textit{e.g.}, Refs.~\cite{Krall1973,Karniadakis1988}).
But, here, the measured numerical values are larger due to the contribution of the shell that further slows down solute supply to the reservoir surface, and consequently decreases the concentration boundary layer thickness.
$\rm{Sh}(\theta,{\rm T})$ stays steady as long as $c_{\rm s}(\theta,{\rm T})$ and $\varphi(\theta,{\rm T})$ have not yet vanished.
\subsection{Instantaneous average Sherwood number}
\label{subsec:average_sherwood}
\begin{figure}[H]
\centering
\includegraphics[width = 0.5\textwidth]{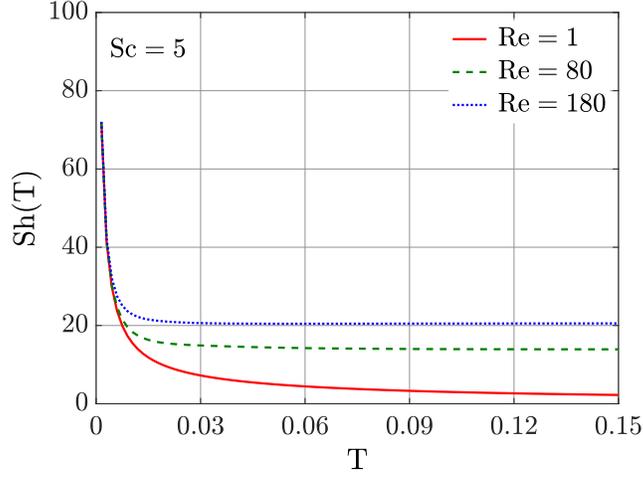}
\caption{Evolution in time of the average Sherwood number $\mathrm{Sh}({\rm T})$ for mass transfer from a cylindrical core-shell reservoir in crossflow at various Reynolds numbers ${\rm Re}$.
$\mathrm{Sh}({\rm T})$ reaches a plateau at larger Reynolds numbers.
Here, ${\rm Sc} = 5$ and ${\rm P} = 3.70$.
}
\label{fig:average_sherwood}
\end{figure}

The instantaneous average Sherwood number $\rm{Sh}({\rm T})$, evaluated with Eq.~(\ref{eq:average_sh}), is reported in Fig.~(\ref{fig:average_sherwood}) for different flow regimes.
$\rm{Sh}({\rm T})$ is transiently infinite, at the early stage of all the simulations, when the solute still diffuses through the shell and has not yet come into contact with the fluid.
Later on the three curves diverge from each other when the solute manages to reach the surface of the reservoir.
For $\rm{Re} = 1$, $\rm{Sh}({\rm T})$ decays continuously towards zero since the concentration boundary layer expands unsteadily as the solute is released. 
However, at larger Reynolds numbers, the released solute is quickly moved away from the reservoir surface by the flow, and this causes the concentration boundary layer to shrink into a steady thin thickness. 
This is why the average Sherwood number adopts a steady plateau, whose value $\rm{Sh}$ increases with $\rm{Re}$, \textit{i.e.}, when the mass transfer is further enhanced by forced convection.
$\mathrm{Sh}({\rm T})$ reaches quickly its steady values as ${\rm Re}$ is increased.
The transient time depends on the problem parameters.
The unsteady and the oscillatory characters of the von K\'{a}rm\'{a}n vortex street has no visible qualitative signature on the overall temporal evolution of the average Sherwood number.
Fluctuations may show up if ${\rm Re}$ is further increased to trigger turbulent flows, as reported for a cylinder without a shell in Ref.~\cite{Bouhairi2007}.
The emergence of the recirculations or vortex shedding, at the rear of the reservoir, influences only the numerical value adopted by $\mathrm{Sh}({\rm T})$.
The dependency of the steady value of $\mathrm{Sh}({\rm T})$ on other factors, such as the Schmidt number ${\rm Sc}$ and the dimensionless shell permeability to solute ${\rm P}$ is investigated in the following.
\subsection{Solute release kinetics}
\label{subsec:mass_release}
\begin{figure}[H]
\subfloat[\label{subfig:release_vs_T}]{\includegraphics[width=0.45\textwidth]{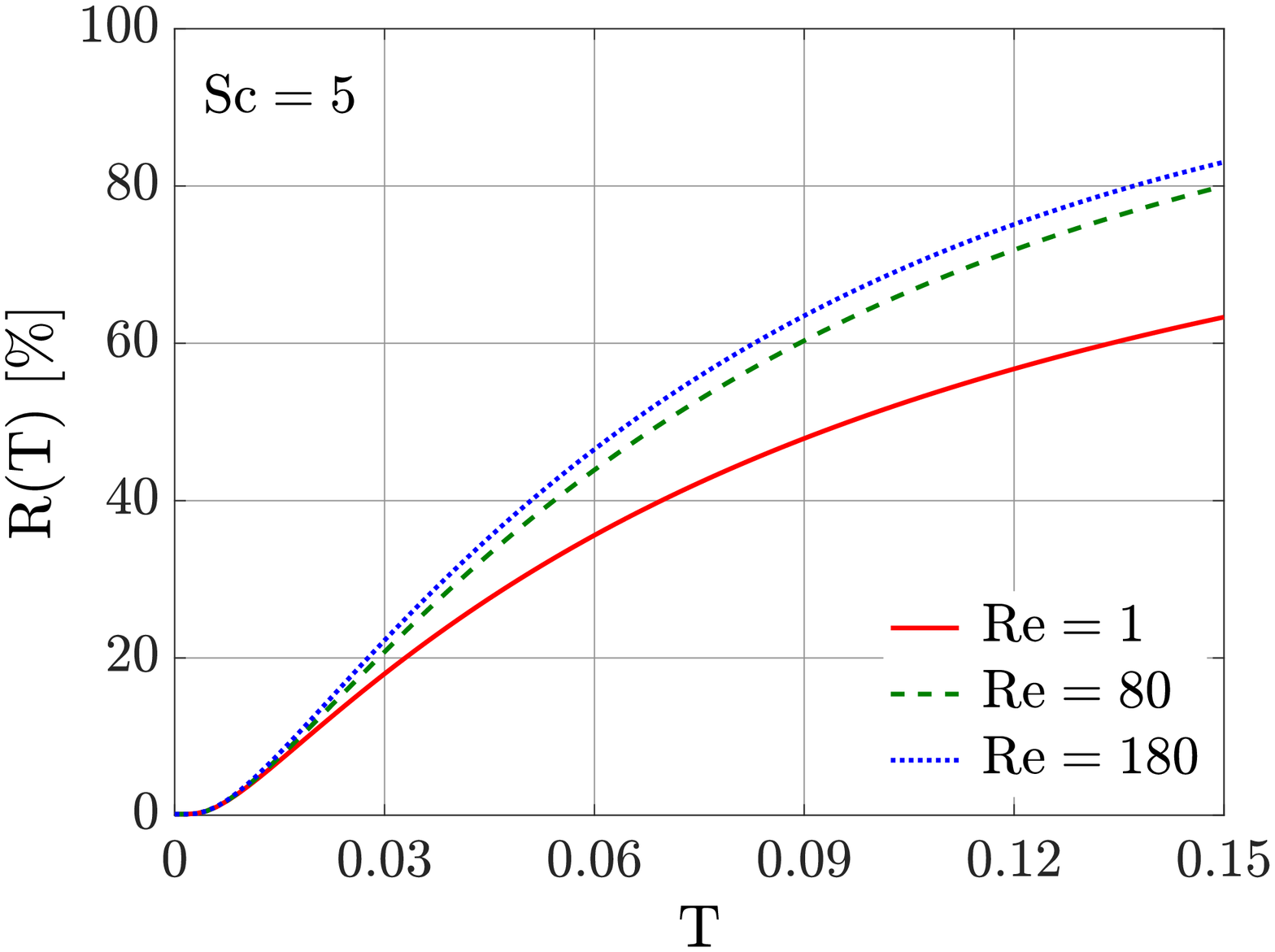}}
\hfill
\subfloat[\label{subfig:release_vs_sqrt(T)}]{\includegraphics[width=0.45\textwidth]{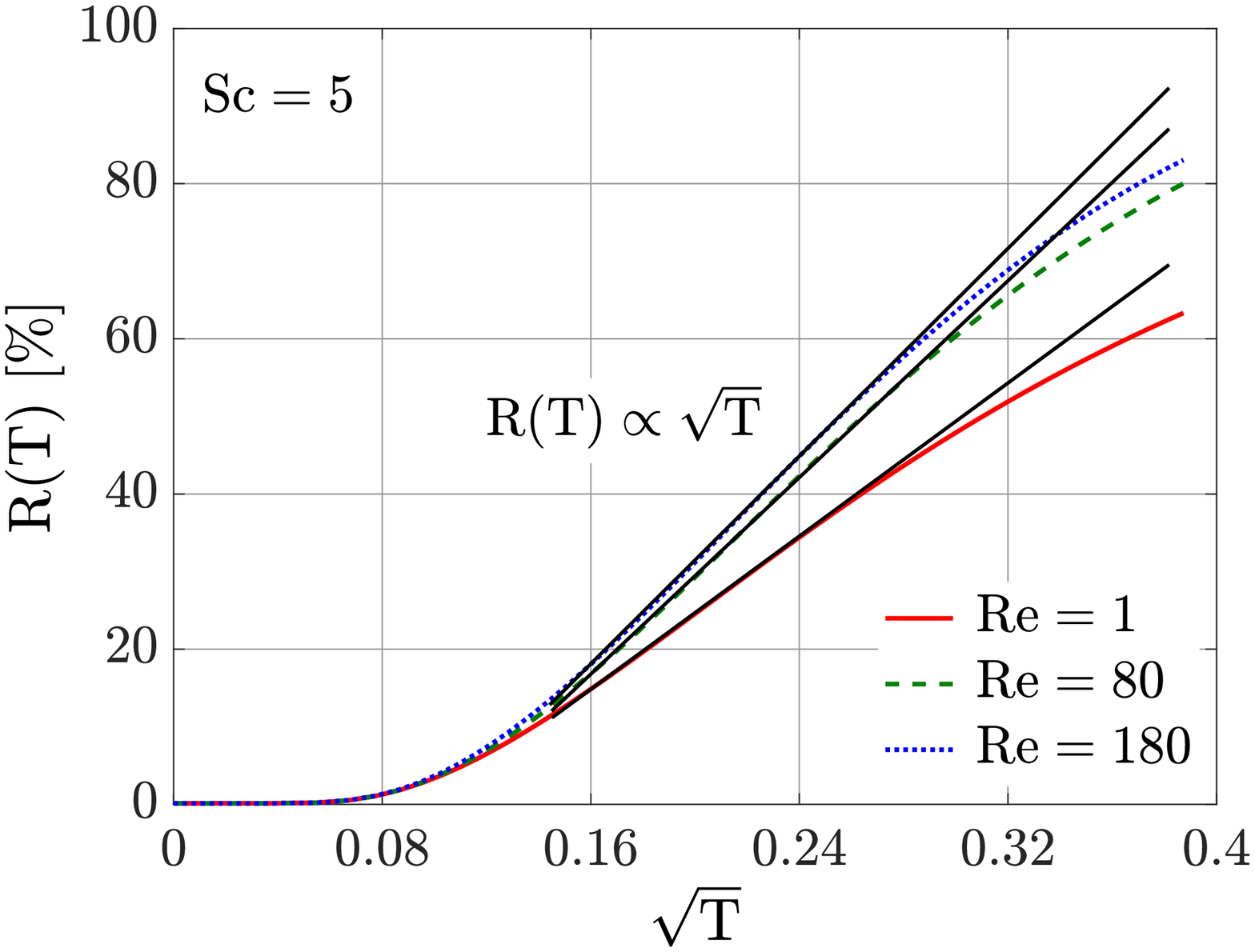}}  
\caption{Release rate of the solute from the cylindrical core-shell reservoir ${\rm R}({\rm T})$ as a function of time $\mathrm{\rm T}$ (a), and $\sqrt{\mathrm{\rm T}}$ (b) at various Reynolds numbers ${\rm Re}$.
(b) shows also a comparison with the Higuchi model ${\rm R}({\rm T}) = a\sqrt{\rm T}$ largely used in pharmacokinetics \cite{Higuchi,Siepmann2012}.
${\rm R}({\rm T})$ is enhanced when increasing ${\rm Re}$ due to forced convection.}
\label{fig:mass_release}
\end{figure}

Another way of quantifying mass transfer from a reservoir, at the macroscopic scale, is the solute release rate over time $\rm{R}({\rm T})$. 
This is largely used to characterize drug release from particles in the pharmacokinetics \cite{Siepmann2012}, and it is measured as follows:
\begin{equation}
\mathrm{R}({\rm T}) = \frac{\text{Instantaneous released solute into the fluid}}{\text{Initially loaded solute in the reservoir}}= 100\times\left(1-\frac{m({\rm T})}{m(0)}\right)\text{,}
\end{equation}
where $m({\rm T})$ is the total mass of the solute that remains inside the reservoir at dimensionless time ${\rm T}$ \cite{Kaoui2018}:
\begin{equation}
\qquad m({\rm T}) = \int_{-\frac{W}{2}}^{+\frac{W}{2}}\int_{-\frac{L}{2}}^{+\frac{L}{2}} c(x,y,{\rm T})I(x,y)dxdy\text{,}
\end{equation}
where $I(x,y) = 1$ inside the reservoir and $I(x,y) = 0$ elsewhere.
Figure \ref{fig:mass_release} gives the evolution in time of the total mass of the released solute from the core-shell reservoir $\mathrm{R}({\rm T})$ for three flow regimes when ${\rm Sc} = 5$ and ${\rm P} = 3.70$. 
Initially, the whole solute is encapsulated in the core. 
Afterwards, the solute diffuses from the core to the surrounding fluid through the shell.
During this pure diffusion process within the shell, the flow has no impact yet on the release rate and the evolution is similar for the three curves.
As soon as the solute reaches the reservoir surface, it starts being advected by the flow. 
This explains why the curves detach from each other from that moment on.
During the second stage (${\rm T} > 0.015$), the release rate increases rapidly. 
It evolves as the square root of the time for a short lapse of time (see Fig.~\ref{subfig:release_vs_sqrt(T)}).
This follows the Higuchi model, ${\rm R}({\rm T}) = a\sqrt{{\rm T}}$ \cite{Higuchi}, which is originally used to predict drug release from particles in absence of external flow, \textit{i.e.}, under static conditions \cite{Siepmann2012}.
Here, this model approximates well the computed data, at short-term, even though the core-shell is subjected to an external flow.
At long-term, however, the release rate deviates from the square root model when the release slows down gradually as the reservoir exhausts its solute.
The release rate is faster at larger ${\rm Re}$ due to forced convection that enhances the solute delivery efficiency.
\subsection{Effect of the shell solute permeability}
The main shell property of interest in this study is its permeability to the solute $p_{\rm s}$ (the shell mass transfer coefficient), which must not be confused with the solvent permeability of the shell that characterizes rather the transport of the solvent through the shell.
Here, the core-shell reservoir is considered to be nonporous to the fluid.
Understanding and quantifying the effect of the shell solute permeability on mass transfer is essential in designing efficient controlled delivery systems, and devices that use core-shell or hollow fibers to control mass transfer rate.
In this study, the value of the dimensionless shell solute permeability ${\rm P}$ (Eq.~(\ref{eq:scaled_p})) is varied by tuning solely $D_{\rm s}$ in such way to achieve $0 < D_{\rm s}/D_{\rm f} \leq 1$.
This allows to examine the effect of ${\rm P}$, while holding the same geometrical setup of the problem and the same flow pattern around the cylinder, in contrast if ${\rm P}$ had been varied through $\delta$.
When ${\rm P}$ tends to zero, it means the shell becomes impermeable to the solute.

\begin{figure}[H]
\centering
\label{subfig:concentration_profiles_tau_s_0_53}{\includegraphics[width=0.45\textwidth]{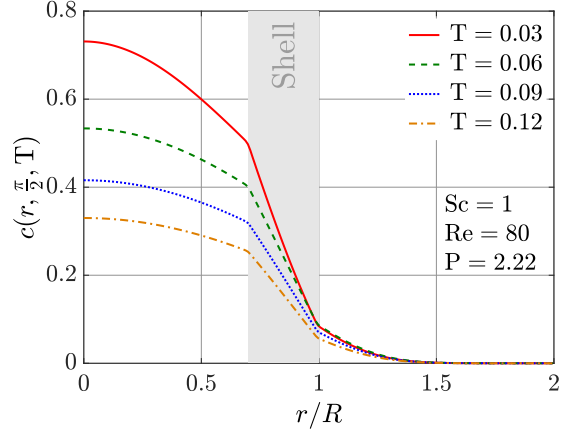}}    
\caption{Solute concentration profiles across the shell of the reservoir for a lower shell solute permeability ${\rm P} = 2.22$.
The location of the shell is materialized with the gray-colored area.
The solute has the same diffusion coefficient in both the core (left side of the shell) and the fluid (right side of the shell) $D_{\rm c}=D_{\rm f}$, but it has a smaller coefficient in the shell $D_{\rm s} < D_{\rm f}$.
This leads to the emergence of a jump in the solute concentration across the shell, and to the slowdown of the solute release.}
\label{fig:concentration_profiles}
\end{figure}

\begin{figure}[H]
\centering
\subfloat[\label{subfig:Sh_eq_vs_P}]{\includegraphics[width=0.45\textwidth]{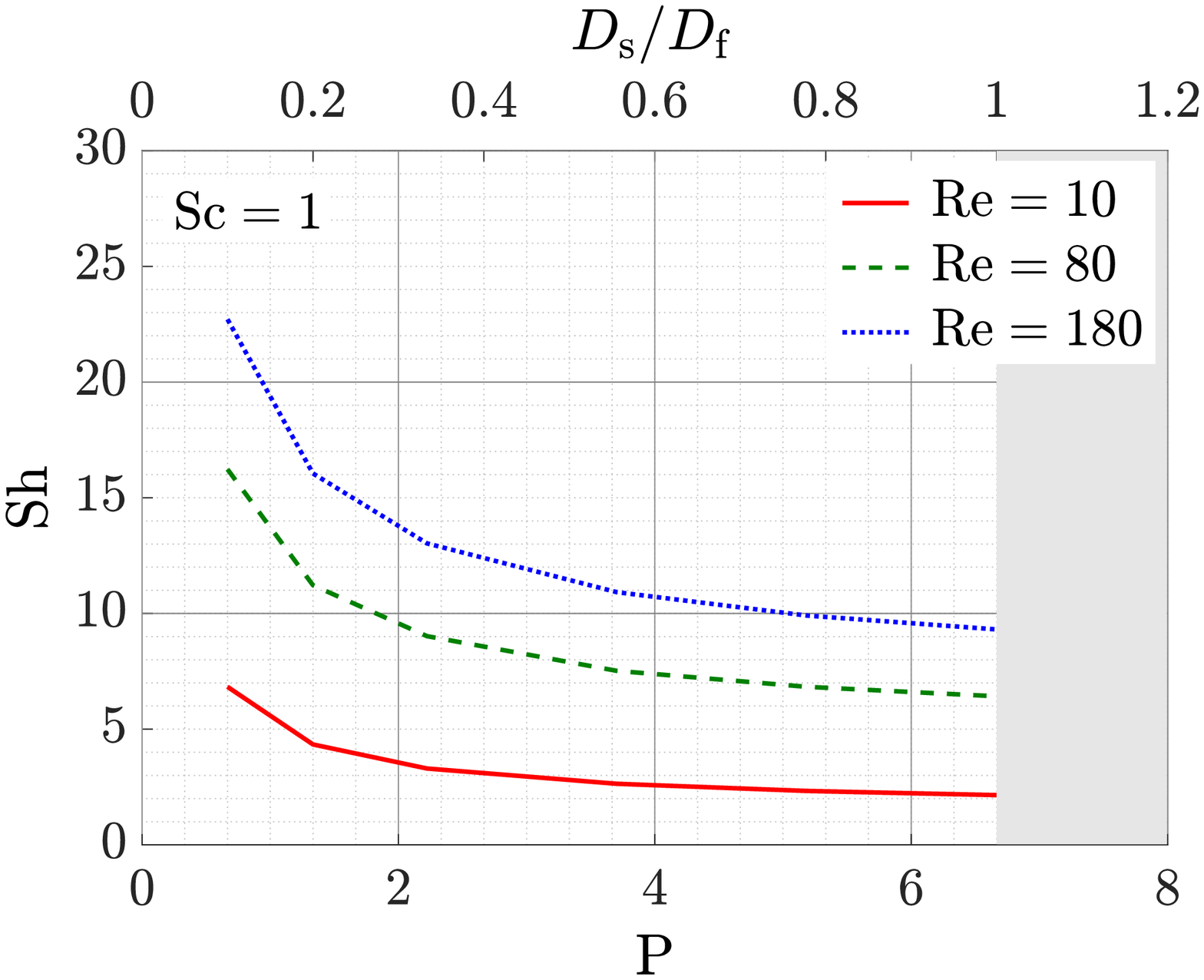}}  
\hfill
\subfloat[\label{subfig:Sh_eq_vs_P_loglog}]{\includegraphics[width=0.45\textwidth]{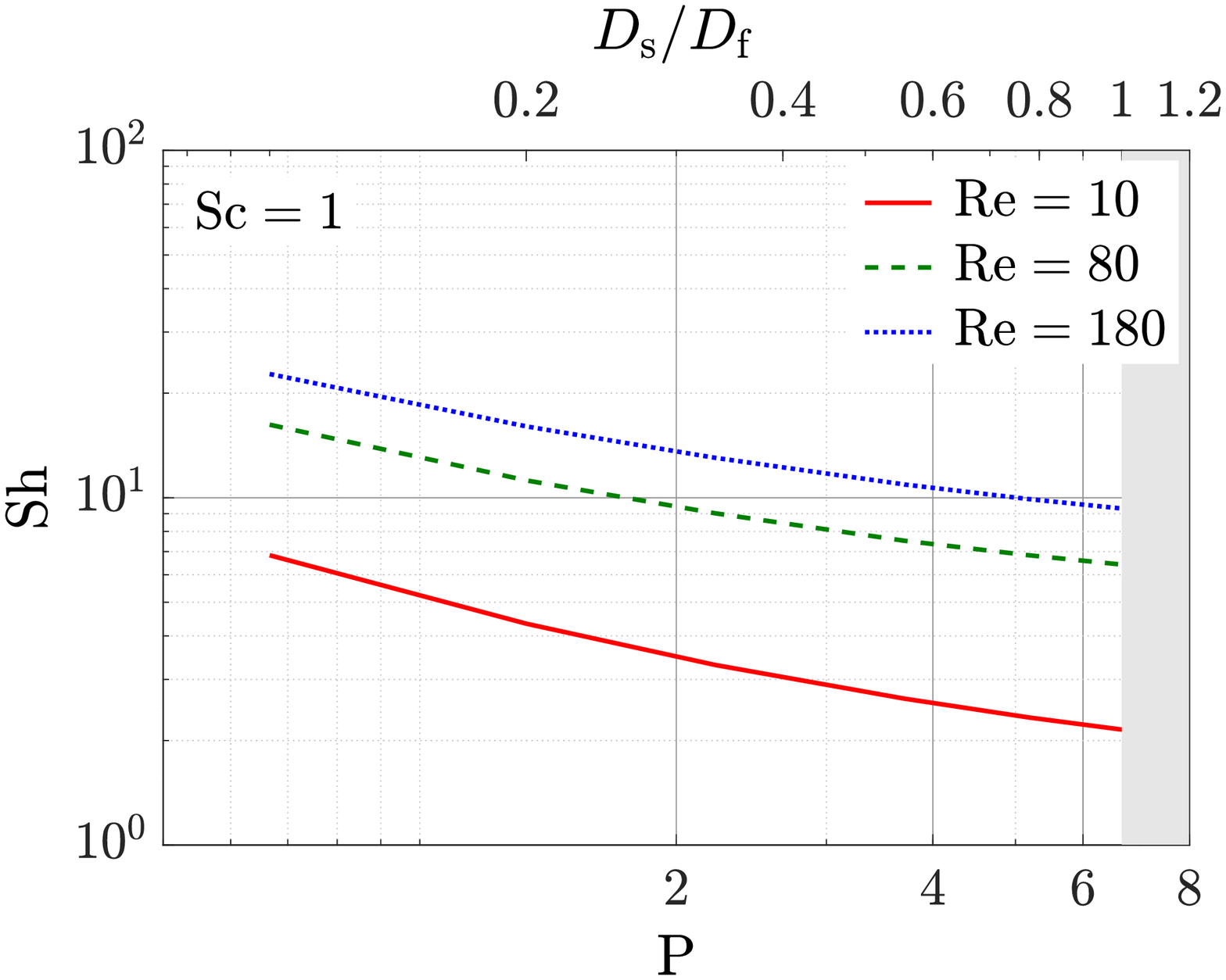}}
\caption{Steady average Sherwood number ${\rm Sh}$ of mass transfer from a core-shell cylinder in cross flow, as a function of the dimensionless shell solute permeability ${\rm P}$ and the shell to fluid diffusion coefficients ratio $D_{\rm s}/D_{\rm f}$, at different flow regimes. 
The Schmidt number is ${\rm Sc}=1$.
(a) and (b) show data plotted in linear and log-log scales, respectively.
${\rm Sh}$ decreases as ${\rm P}$ increases.
The grey area limit is not studied since it corresponds to the nonphysical situation of $D_{\rm s}>D_{\rm f}$.}
\label{fig:Sh_P}
\end{figure}

The effect of a lower shell solute permeability, ${\rm P}=2.22$, on the concentration profiles is shown in Fig.~\ref{fig:concentration_profiles}.
These profiles are taken at $x=x_c$ ($\theta = \pi/2$) and at different dimensionless times.
The concentration profiles tend to adopt linear variation within the shell, with decaying slopes as the solute is leaking out of the reservoir.
The shell acts as a barrier that slows down the release rate of the solute.
Shells with lower values of ${\rm P}$ slow down delivery of solute to the reservoir surface, which
lowers the contribution of diffusion to the mass transfer rate at the adjacent flowing fluid.
The influence of the dimensionless shell solute permeability ${\rm P}$ on the steady average Sherwood number $\rm{Sh}$, at different flow regimes, is reported in Fig.~\ref{fig:Sh_P} with both linear and log-log scales.
The results are also shown as a function of the diffusivity ratio $D_{\rm s}/D_{\rm f}$, with $D_{\rm f} = D_{\rm c}$.
Each curve corresponds to a specific flow regime: steady laminar ($\mathrm{Re} = 10$), steady laminar with recirculations ($\mathrm{Re} = 80$) and unsteady with vortex shedding ($\mathrm{Re} = 180$). 
$\rm{Sh}$ decreases in both linear and log-log scales for all the flow regimes.
Figure~\ref{subfig:Sh_eq_vs_P} shows deep decline of $\rm{Sh}$ at lower values of ${\rm P}$.
These results reveal for the first time the effect of the shell solute permeability on mass transfer from a circular cylinder, coated with a semi-permable shell, under crossflow.
The present study highlights how the internal structure, of a cylinder, and the transport properties of its shell alter substantially the mass transfer performance.
While one would expect enhancement of ${\rm Sh}$ by increasing ${\rm P}$ because this later enhances the mass transfer rate ${\rm R}({\rm T})$ across the shell, data in Fig.~\ref{subfig:Sh_eq_vs_P} show the opposite scenario.
For a given flow pattern, increasing ${\rm P}$, while holding all other parameters constant, enhances solute release ${\rm R}({\rm T})$ that further expands the thickness of the concentration boundary layer, to which ${\rm Sh}$ is inversely proportional.
This is why ${\rm Sh}$ decreases with increasing ${\rm P}$.
\subsection{Correlation for mass transfer coefficient}
\label{subsec:correlation}
A correlation for mass transfer coefficient that takes into consideration the effect of the shell solute permeability is explored.
Mass transfer correlations are usually obtained by scaling experimental data of the steady average Sherwood number ${\rm Sh}$ as a function of mainly the Reynolds number ${\rm Re}$ and the Schmidt number ${\rm Sc}$.
A general form has been proposed by Hilpert for a single circular cylinder in crossflow \cite{Hilpert1933}, 
\begin{equation}
\mathrm{Sh} = c\mathrm{Re}^{m}\mathrm{Sc}^n.
\label{eq:hilpert}
\end{equation}
This correlation is adapted from the heat transfer problem to the actual mass transfer problem by replacing the Nusselt number $(\mathrm{Nu})$ by the Sherwood number $(\mathrm{Sh})$, and the Prandtl number $(\mathrm{Pr})$ by the Schmidt number $(\mathrm{Sc})$, which is possible in the limit of negligible viscous dissipation.
The prefactor $c$ depends on ${\rm Re}$ and on other factors, such as the shape of the cylinders.
The value of the power exponent $m$ depends also on ${\rm Re}$, and it is generally within the range of $0.3 \leq m \leq 0.5$ in absence of turbulence \cite{Hilpert1933,Incropera2017}. 
Most studies adopts $n = \frac{1}{3}$, which is evaluated by the classical boundary layer theory \cite{Leveque1928}; however, in some case studies, deviation from $\frac{1}{3}$ is measured \cite{Zukauskas1972}.
There are many other correlations, and they come with different degrees of accuracy \cite{Zukauskas1972,Whitaker1972,Morgan1975}.
The most popular one is the correlation of Churchill-Bernstein \cite{Churchill1977},
\begin{equation}
{\rm Sh} = 0.3 + 
\frac{{0.62{\rm Re}^{1/2}}{\rm Sc}^{1/3}}
{\left[1 + \left(0.4/{\rm Sc}\right)^{2/3} \right]^{1/4}}\left[1 + \left(\frac{{\rm Re}}{282,000}\right)^{5/8} \right]^{4/5},
\end{equation}
that, despite its complexity compared to Hilpert's correlation, has the advantage of being valid for a wide range of the P\'{e}clet number  ${\rm Re}\,{\rm Sc} \geq 0.2$, and without involving exponents that need to be adjusted depending on either ${\rm Re}$ or ${\rm Sc}$.
This correlation, and all others in literature, consider Dirichlet or Neumann boundary conditions, which correspond to either constant concentration or constant mass flux at the surface of the cylinder.
Thus, they are not adapted to the case of a cylindrical reservoir that implies unsteady continuous boundary conditions (Eqs.~\ref{eq:bc_cs} and \ref{eq:bc_sf}).
Moreover, they do not consider the presence of a coating semipermeable shell.
In this study, the general form of Hilpert, Eq.~(\ref{eq:hilpert}), is used and the dependency of its power exponents and prefactor on ${\rm Re}$, ${\rm Sc}$, and most importantly on ${\rm P}$ is examined.
\begin{figure}[H]
\centering
\includegraphics[width = 0.45\textwidth]{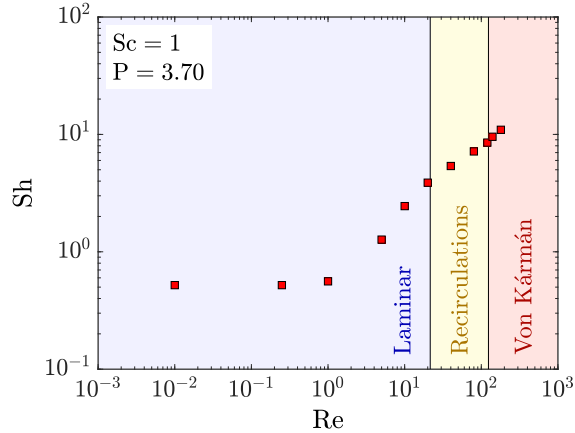}
\caption{Steady average Sherwood number ${\rm Sh}$ versus Reynolds number ${\rm Re}$ at different flow regimes.
The Schmidt number is $\mathrm{Sc} = 1$ and the dimensionless shell solute permeability is ${\rm P}=3.70$.
${\rm Sh}$ is insensitive to ${\rm Re}$ when ${\rm Re} \leq 1$, while it increases at larger ${\rm Re}$ due to the contribution of forced convection.}
\label{fig:sherwood_regimes}
\end{figure}

Figure \ref{fig:sherwood_regimes} reports the steady values of the average Sherwood number $\mathrm{Sh}$ versus the Reynolds number ${\rm Re}$ in a log-log plot for a core-shell cylinder with dimensionless shell solute permeability ${\rm P}=3.70$, and at $\mathrm{Sc} = 1$.
Data are obtained within the range $0.01 \leq {\rm Re} \leq 180$ that covers three flow regimes.
Two main behaviors are obtained depending on the value of ${\rm Re}$. 
For the Stokes flow regime ($\mathrm{Re} < 1$), $\mathrm{Sh}({\rm T})$ reaches a quasi-steady regime (see Fig.~\ref{fig:average_sherwood}) and the reported values of $\mathrm{Sh}$ in Fig.~\ref{fig:sherwood_regimes} are taken at ${\rm T}=0.15$.
In this regime, $\mathrm{Sh}$ is independent of ${\rm Re}$ and it is relatively low since the mass transfer is rather dominated by diffusion. 
$\mathrm{Sh}$ declines very slowly toward zero as the solute is released from the reservoir and the concentration boundary layer continues to expand without adopting a steady thickness (Fig.~\ref{fig:flow_mass_transfer_Re_1}).
At long term, ${\rm Sh}$ is expected to vanish as indicated in Ref.~\cite{Whitaker1972} (${\rm Sh}\rightarrow 0$ as ${\rm Re}\rightarrow 0$).
For $\mathrm{Re} \gg 1$, advection becomes the dominant mass transfer mechanism, and consequently ${\rm Sh}$ increases linearly in the log-log scale.
In the range $1 \leq \mathrm{Re} \leq 10$, the flow regime transits from Stokes flow to separated flow, and the mass transfer is operated equivalently by both advection and diffusion, with none of them being predominant.
Figure \ref{fig:sherwood_regimes} shows similar main qualitative behaviors and features as reported for other particles and obstacles \cite{Clift1978,Michaelides2006,Incropera2017}, but with different numerical values due to the contribution of the shell and the unsteady boundary conditions.

\begin{figure}[H]
\subfloat[\label{subfig:Sh_eq_vs_Re}]{\includegraphics[width=0.45\textwidth]{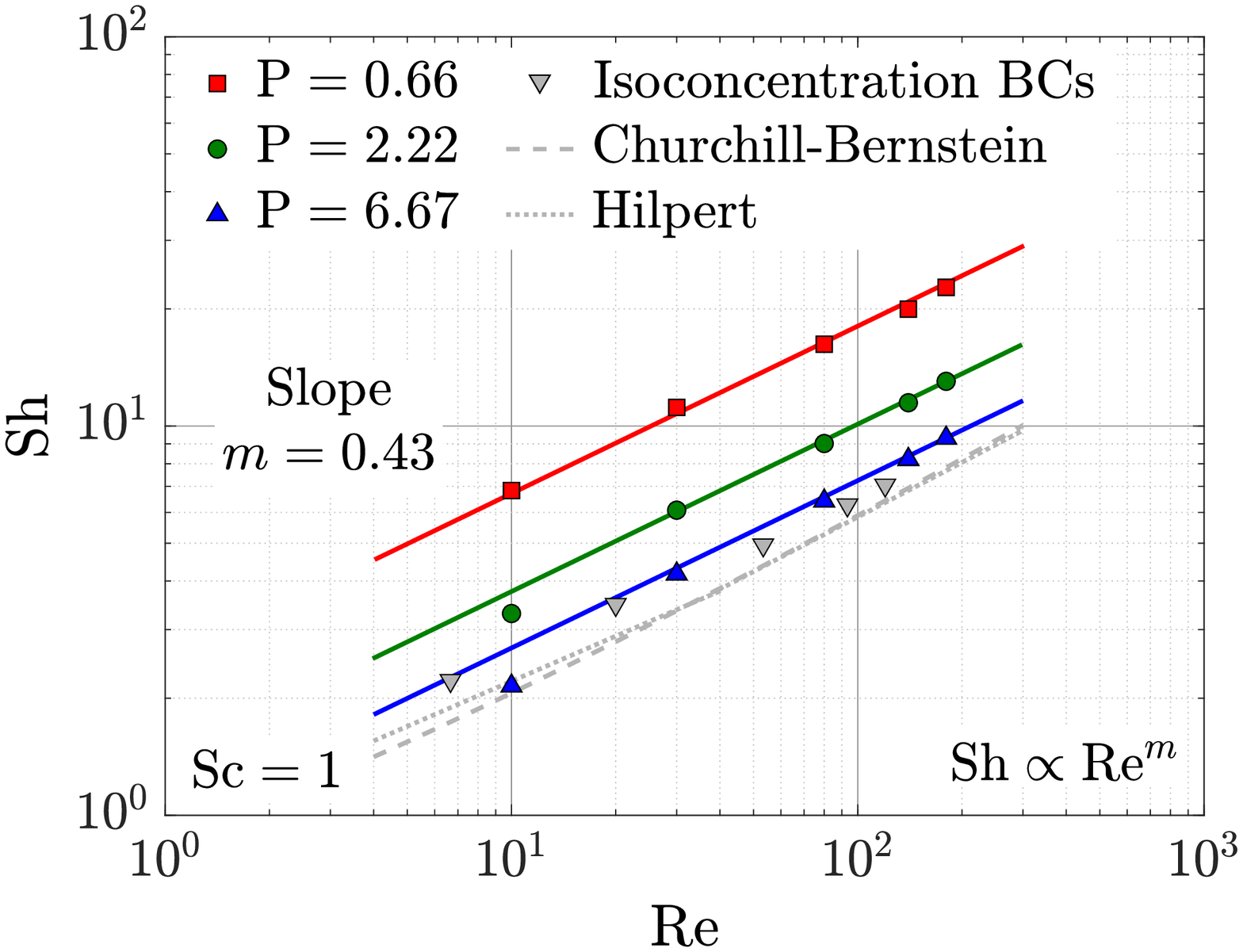}}  
\hfill
\subfloat[\label{subfig:Sh_eq_vs_Sc}]{\includegraphics[width=0.45\textwidth]{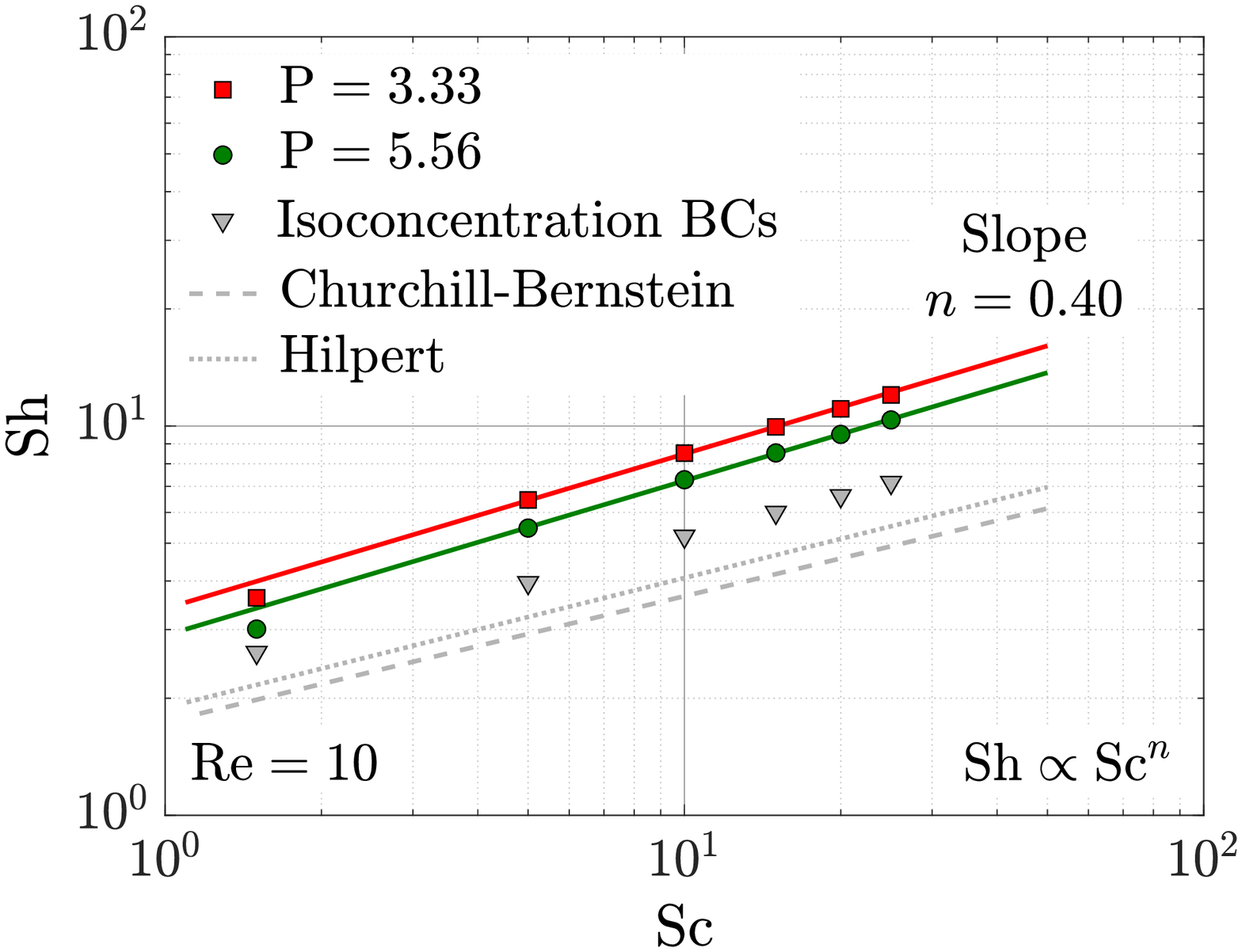}}
\caption{
(a) Sherwood number ${\rm Sh}$~\textit{vs.}~the Reynolds number ${\rm Re}$, and (b) Sherwood number ${\rm Sh}$~\textit{vs.}~the Schmidt number ${\rm Sc}$ for different dimensionless shell solute permeabilities ${\rm P}$.
The computed data are fit linearly to extract the correlation power exponents $m=0.43$ and $n=0.40$, which visibly do not depend on ${\rm Re}$ and ${\rm Sc}$.
For comparison, the equations of Churchill-Bernstein \cite{Churchill1977} and Hilpert \cite{Hilpert1933,Incropera2017} are plotted, and data points from simulations with constant surface concentration on a cylinder without a shell are also shown.
The presence of the shell shifts ${\rm Sh}$ upwards to larger values.
}
\label{fig:correlation_loglog}
\end{figure}
\begin{figure}[H]
\centering
\includegraphics[width = 0.45\textwidth]{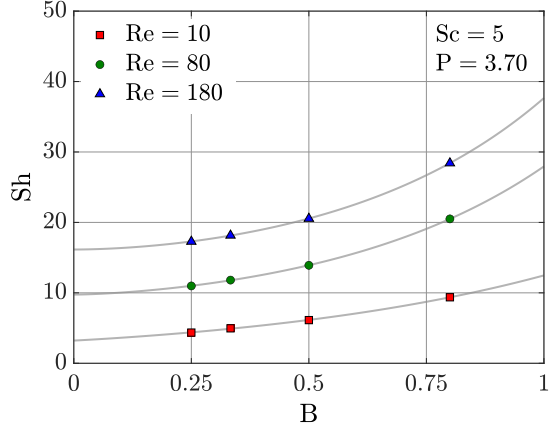}
\caption{${\rm Sh}$ as a function of the blockage ratio ${\rm B}$ for a cylindrical core-shell reservoir, with dimensionless shell solute permeability ${\rm P}=3.7$, in crossflow at various ${\rm Re}$ and at ${\rm Sc}=5$.
The grey solid lines are fit of the numerical data with a mathematical formula of the form $\alpha \exp \left( \beta {\rm B} ^{\gamma}\right)$, where the parameters depend on ${\rm Re}$. Extrapolation of the data with this function gives the expected values of ${\rm Sh}$ at ${\rm B}=0$ (unconfined cylindrical  reservoir).}
\label{fig:confinement}
\end{figure}

The contribution of the shell to the mass transfer correlation is analyzed only for large Reynolds numbers $10 \leq {\rm Re} \leq 180$, upon which $\mathrm{Sh}$ depends strongly.
$\mathrm{Sh}$ is plotted as a function of ${\rm Re}$ and ${\rm Sc}$ in log-log scale for different dimensionless shell solute permeabilities ${\rm P}$ in Fig.~\ref{fig:correlation_loglog}. 
${\rm Sc}$ is varied from $1$ to $25$.
The numerical data are represented with symbol points, and their respective fit with solid lines. 
For comparison, numerical data obtained for a naked cylinder, without a shell, and with constant surface concentration, \textit{i.e.}, isoconcentation boundary condition, are also reported (down-pointing triangles).
For these data, the Reynolds number is defined as ${\rm Re}=U_\infty d/\nu$ to allow quantitative comparison with the classical correlations of Churchill-Bernstein \cite{Churchill1977} (dashed line) and Hilpert \cite{Hilpert1933} (dotted line), which are also plotted in Fig.~\ref{fig:correlation_loglog}.
For the core-shell cylinder, the fits of the numerical data in both subfigures are linear and adopt the same slope in each subfigure regardless of the value of ${\rm P}$.
The measured slopes do not depend on either ${\rm Re}$ or ${\rm Sc}$.
${\rm Sh}$ increases with decreasing ${\rm P}$, as explained in the previous section.
The few data points that deviate from the solid lines at either low ${\rm Re}$ or ${\rm Sc}$, or large ${\rm P}$ are discarded in the analysis.
For these parameters, the advection is relatively weak; and the continuous expansion of the concentration boundary layer is altered by the channel confining walls.
All the data computed for the core-shell reservoir are larger than those computed for a cylinder without a shell and with constant surface concentration.
Even the largest used ${\rm P}$, whose case corresponds to a reservoir with an internal homogeneous structure ($D_{\rm s} = D_{\rm c}$), leads to larger values of ${\rm Sh}$.
The correlations of Hilpert \cite{Hilpert1933} and Churchill-Bernstein \cite{Churchill1977} both fall below all other data since they are obtained for unconfined cylinders (${\rm B}=0$).
They are located just below the numerical data for constant surface concentration, which are obtained for a confined cylinder ${\rm B}=0.5$.
The increase of the blockage ratio ${\rm B}$ from $0$ to $0.5$ increases ${\rm Sh}$, but not as much as the presence of a shell does.
It is remarkable how decreasing ${\rm P}$ shifts ${\rm Sh}$ upwards far from the values of the classical correlations. 
Figure \ref{fig:correlation_loglog} suggests a correlation of the form:
\begin{equation}
{\rm Sh} = c{\rm Re^{0.43}}{\rm Sc^{0.40}},
\label{eq:correlation1}
\end{equation}
where the prefactor $c$ is certainly a function of the shell solute permeability ${\rm P}$.
Other couple of exponents than $(0.43,0.40)$ have been tested to fit the data, but all of them lead to larger errors.
The Reynolds number exponent $m = 0.43$ is within the range $\frac{1}{3} \leq m \leq \frac{1}{2}$ that is consistent with Hilpert's correlation for the studied range of the Reynolds numbers (${\rm Re} < 4000$).
The deviation of the Schmidt number exponent $n$, from its classical value of $1/3$, could be explained by the type of the boundary condition on the reservoir, which is unsteady and continuous; and thus, differs from the constant concentration and constant mass flux boundary conditions previously used to derive analytically $n=\frac{1}{3}$.
The deviation could be also attributed to the effect of the wall confinement since the cylindrical reservoir is confined throughout this study.
Indeed, even when setting constant concentration boundary condition on the cylinder, but with ${\rm B}=0.5$, $n=0.35$ is measured for data shown with down-pointing triangles in Fig.~\ref{fig:correlation_loglog}(b), which is also slightly larger than $1/3$.
\begin{figure}[H]
\centering
\includegraphics[width = 0.45\textwidth]{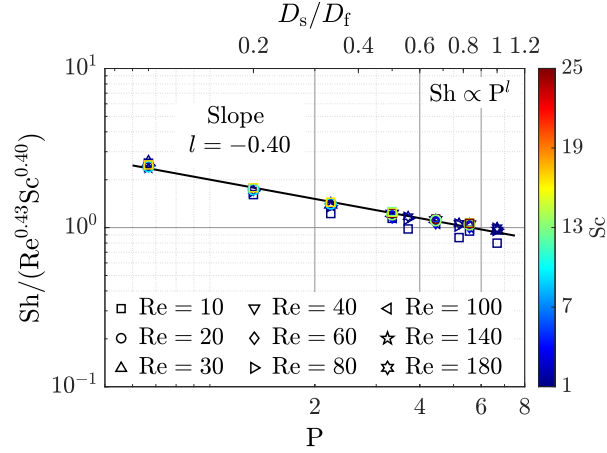}
\caption{Log-log plot of the scaled steady average Sherwood number ${\rm Sh}$, \textit{i.e.}, the correlation prefactor $c$ in Eq.~(\ref{eq:correlation1}), versus the dimensionless shell solute permeability ${\rm P}$, with its corresponding ratio of the solute diffusivities in the shell and in the fluid $D_{\rm s}/D_{\rm f}$, for mass transfer from a cylindrical core-shell reservoir in crossflow at various combinations of the Reynolds number ($10 \leq \mathrm{Re} \leq 180$) and the Schmidt number ($1 \leq \mathrm{Sc} \leq 25$). Most data points collapse into a line with a slope of $l=-0.40$.}
\label{fig:scatter_plot}
\end{figure}
It remains to determine how $c$ depends on ${\rm P}$ in order to highlight the effect of the shell.
In the following, this is determined for the blockage ratio ${\rm B}=0.5$, which is set along the present study.
The slight deviation of data computed for the case of constant concentration boundary condition on a cylinder without any shell with respect to the correlations of Hilpert and Churchill-Bernstein, observed in Fig.~\ref{fig:correlation_loglog}, suggests that the effect of confinement is not as such significant at the blockage ratio ${\rm B}=0.5$, as also discussed in Ref.~\cite{Zukauskas1972}.
Figure \ref{fig:scatter_plot} gives the scaled correlation prefactor $c={\rm Sh}/({\rm Re^{0.43}}{\rm Sc^{0.40}})$ versus the shell solute permeability ${\rm P}$, and the corresponding ratio of the diffusivity in the shell to the one in the fluid $D_{\rm s}/D_{\rm f}$.
The numerical data are obtained within the Reynolds number range of $10 \leq \mathrm{Re} \leq 180$ and the Schmidt number range of $1 \leq \mathrm{Sc} \leq 25$.
All data collapse into a single line of slope $l=-0.40$, except few that correspond to low Péclet numbers $\mathrm{Re}\,\mathrm{Sc}$.
Figure \ref{fig:scatter_plot} suggests a correlation of the form,
\begin{equation}
{\rm Sh} = 1.99{\rm P}^{-0.40}{\rm Re^{0.43}}{\rm Sc^{0.40}},
\label{eq:proposed_correlation}
\end{equation}
to estimate mass transfer from a single cylindrical core-shell reservoir placed in a crossflow. 
Equation~(\ref{eq:proposed_correlation}) correlates well all the numerical data with an error of $4\%$.
Here, independently of the effect of the blockage, the proposed correlation Eq.~(\ref{eq:proposed_correlation}) demonstrates the nonnegligible contribution of the presence of a coating semipermeable shell on the mass transfer coefficient.
It does also highlight the effect of having an internal nonhomogeneous structure of a cylinder.
Equation~(\ref{eq:proposed_correlation}) has been obtained by screening a wide range of parameter space by varying ${\rm P},$ ${\rm Re}$ and ${\rm Sc}$ ($100$ high-resolution and accurate simulations in total have been carried out for this study).
Further systematic analysis of the dependency of $c$, and even $m$ and $n$, on the blockage is essential before proposing a universal correlation.
This is left for a future investigation. 
Here, the effect of the blockage on mass transfer coefficient is briefly explored for a given value of the dimensionless shell solute permeability ${\rm P}=3.7$, and at ${\rm Sc}=5$, see Fig.~\ref{fig:confinement}.
${\rm Sh}$ increases with increasing ${\rm B}$, \textit{i.e.}, by confining further the cylindrical  reservoir between the channel walls.
This agrees with the behavior of naked cylinders whose local and global heat transfer are found to be influenced by the blockage as mentioned earlier in Refs.~\cite{Zukauskas1972,Morgan1975,Khan2004}.
The confinement speeds up the fluid flow within the gap between the cylinder surface and the wall, which amplifies the contribution of the convection; and hence, it increases ${\rm Sh}$.
The data points are fit perfectly with a mathematical function of the form $\alpha \exp \left( \beta {\rm B} ^{\gamma}\right)$, where the fit parameters $\alpha$, $\beta$ and $\gamma$ depend on ${\rm Re}$.
Estimation of ${\rm Sh}$ at ${\rm B}=0$ is obtained by extrapolation.
Figure~\ref{fig:confinement} shows that ${\rm Sh}$ varies slowly within the range $0 \leq {\rm B} \leq 0.5$, and it varies very much when ${\rm B} > 0.5$ that corresponds to the strong confinement limit.
Moreover, the estimated values at ${\rm B}=0$ are larger than the values predicted by the correlation of Hilpert (Eq.~(\ref{eq:hilpert})) due to the presence of the coating shell.
\section{Conclusions}
\label{sec:conclusion}
Mass transfer from a single cylindrical core-shell reservoir in crossflow is studied numerically using two-dimensional simulations based on two-component lattice Boltzmann method.
The main outcome is that the reservoir character leads to unsteady and nonuniform distribution of both the concentration and the mass flux at the cylinder surface.
This mechanism differs from the classical studies, where either concentration or mass flux is set constant (as a control parameter) and the other quantity evolves to steady nonuniform distribution. 
However, the resulting Sherwood number (the dimensionless mass transfer coefficient) manifests the same qualitative, but not quantitative, spatial distribution as if the surface concentration or mass flux has been set constant. 
Thus, having a reservoir - without the shell or when the shell solute permeability is larger - leads to almost similar numerical values as previously reported Sherwood numbers.
Whereas, when the cylindrical reservoir is coated with a shell whose solute permeability tends to zero (\textit{i.e.}, a high mass transfer resistance) the resulting Sherwood number is enhanced.
This is caused by the shell that slows the solute release, and thus, shrinks the thickness of the emerging concentration boundary layer. 
The effect of the semipermeable shell is investigated systematically.
For a given moderate blockage ratio $\mathrm{B}=0.5$, the contribution of the shell is highlighted through its dimensionless solute shell permeability ${\rm P}$ in the proposed correlation, $\mathrm{Sh} = 1.99{\rm  P}^{-0.40}\mathrm{Re}^{0.43}\mathrm{Sc}^{0.40}$, which is extracted from data set computed within the ranges $10 \leq \mathrm{Re} \leq 180$, $1 \leq \mathrm{Sc} \leq 25$, and $0.66 \leq \mathrm{P} \leq 6.67$.
This correlation complements the list of existing mass transfer correlations with the presently studies situation of a cylindrical reservoir coated with a semipermeable shell.
\section*{Acknowledgements }
The authors thank Prof. Howard Stone of Princeton University for his enlightening and helpful inputs, the anonymous referees for their useful comments, and MESRI French ministry and BMBI laboratory for financial support. 
% References

%
\end{document}